\documentclass[traditabstract]{aa} % for the abstract without structuration 
\usepackage{graphicx}
\usepackage{natbib}
\usepackage{txfonts}
\newcommand{\lae}{Ly$\alpha$ }
\newcommand{\lya}{Ly$\alpha$ }

\begin{document}

   \title{Effects of Lyman alpha scattering in the IGM on clustering
     statistics of Lyman alpha emitters}
   \subtitle{}

   \author{C. Behrens
          \inst{1}
          \and
         J. Niemeyer\inst{1}
         \fnmsep
          }
   \institute{Institut f\"ur Astrophysik, Georg-August Universit\"at G\"ottingen,
              Friedrich-Hundt-Platz 1, D-37077 G\"ottingen\\
              \email{cbehren@astro.physik.uni-goettingen.de/niemeyer@astro.physik.uni-goettingen.de}
}

  \abstract
{We investigate the correlations between the observed fraction of \lae
  emission from star-forming galaxies and the large scale structure by
  post-processing snapshots of a large, high resolution hydrodynamical
  simulation with a \lae radiative transfer code at redshifts
  $z=4,\;3,\;2$. We find correlations of the observed fraction with
  density, density gradient along the line of sight, velocity and
  velocity gradient along the line of sight, all within the same order
  of magnitude (tens of percent). Additionally, 
  a correlation with the angular momentum of the dark matter halo is detected. In contrast to a
  previous study, we find no significant deformation of the 2-point
  correlation function due to selection effects
  from radiative transfer in the IGM
  within the limited statistics of the simulation volume.} 


   \maketitle
%
%________________________________________________________________

\section{Introduction}

Galaxies with strong \lae emission features, so-called \lae emitters
(LAEs), are powerful probes of galaxy evolution and cosmological large scale structure.
There are indications that LAEs, or some
 subset thereof, evolved into today's Milky Way type galaxies \citep{Guaita2010a},
 hence their properties may shed light on our own galaxy's
 youth. Also, future observations might be able to detect the first
 galaxies via \lae emission \citep[e.g ][]{Dijkstra2010}.
  
LAEs can be detected very efficiently in narrow-band or integral-field
spectrographic surveys such as the Hobby-Eberly Telescope Dark Energy
Experiment (HETDEX) \citep{Hill2008}. They have been proposed to act as tracers of the underlying matter
distribution at higher redshifts than currently accessible for galaxy
redshift surveys. Specifically, HETDEX aims to use the power spectrum of
$\sim$ 800.000 LAEs to measure the Hubble parameter 
$H(z)$ and angular distance $D_A(z)$ at redshifts between $z=1.9$ and
3.5 with percent-level accuracy in order to constrain the early
dynamics of dark energy. 

Every interpretation of LAE observations needs to take into account
the resonant nature of \lae scattering \citep[e.g. ][]{Cantalupo2005,Dijkstra2006,Adams2009,Zheng2009,Yajima2012a,Schaerer2011a,Hansen2006a}. The large cross section of \lae
photons scattering with neutral hydrogen strongly correlates the
observed \lae spectra and apparent 
luminosities along any given line-of-sight with the density and
velocity structure of the intervening HI \citep{Dijkstra2006,Laursen2011}. As a result, large
amounts of information are encoded in the observations. On scales of
the emitting galaxies and their circumgalactic material, \lae spectra
are very sensitive to the presence of clumps, dust, and in- or
outflows \citep{Zheng2002,Dijkstra2006,Laursen2009,Schaerer2011a,Barnes2011}. Recent simulations also highlight the possibility of
a strong inclination dependence of the \lae observed fraction 
which, in turn, depends on the morphology of the gaseous disk
\citep{Laursen2007b,Yajima2012a,Verhamme2012}. Mapping the theoretical predictions, mostly from numerical
simulations, to properties of observed LAEs has only just begun
\citep[e.g. ][]{NagamineKentaro2010,Shimizu2011a,Forero-Romero2011,Dayal2012} and promises to be a rich field of research in the coming
years. 

On the other hand, correlations of apparent LAE luminosities with the
matter distribution induced by \lae radiation transport (RT) effects
can also contaminate the clustering statistics of LAEs on larger
scales. If they reach out to scales relevant for the extraction of
cosmological parameters, they need to be accounted for by corrections
in the LAE power spectrum in real and redshift space. This effect was 
demonstrated by \cite{Zheng2010} (ZCTM11) using a 
Monte-Carlo \lae RT calculation on the background of a
cosmological simulation snapshot at $z=5.7$. Details of their setup
and further investigations with regards to the luminosity, spectra and
observed fractions of LAEs in their simulation can be found in
\cite{Zheng2009} (ZCTM10) (see also \cite{Zheng2011} for details on extended LAE halos). They found significant
correlations of the \lae observed fraction, i.e.\ the fraction of photons
that are not scattered out of the line-of-sight during their passage
through the intergalactic medium (IGM), with the 
smoothed IGM density and velocity fields. By far the biggest
effect was seen in correlations with the velocity gradient field,
accompanied by a strongly anisotropic signature in the 2-point
correlation function for LAEs in redshift space. If present also at
lower redshifts, an effect of this 
magnitude would seriously affect the interpretation of LAE large-scale
structure surveys like HETDEX. This was investigated in more detail in
\cite{Wyithe2011a} by means of analytic and numerical models for LAE spectra with
in- and outflows, which the authors used to calibrate a modified
parametrization for the LAE power spectrum. Using an Alcock-Paczynski
test, they then showed that the accuracy of HETDEX measurements
could potentially be seriously compromised by \lae RT
effects. In \cite{Greig2012}, this analysis was extended to include the
LAE bispectrum which allows to
break the degeneracy between \lae RT effects and gravitational
redshift-space distortion that is present at the level of the power
spectrum alone.  

While being the most extensive numerical investigation of \lae RT on
cosmological scales, the methodology and resolution of the simulation
analyzed by ZCTM10/11 were inadequate to capture the nonlinear
hydrodynamics in the circumgalactic medium (CGM) surrounding
LAEs. Instead of a full hydrodynamical simulation, ZCTM10/11 employed a
hybrid scheme which assumed hydrostatic equilibrium for the gas in
virialized halos. Consequently, no outflows were present in their
simulation, and the infall was purely gravitational with no
hydrodynamical modifications on CGM scales. In
\cite{Wyithe2011a}, galactic outflows were modeled in a
simplified way that was also assumed to be independent of the
environment on linear scales. The exact degree to which nonlinear flows
on scales $\lesssim 100$ kpc are correlated with their large-scale
environment is still unclear, but can plausibly be assumed to be
non-vanishing. In this case, the well-known strong sensitivity of LAE
properties on CGM/IGM flows \citep[e.g. ][]{Dijkstra2007a,Iliev2008,Laursen2011} will be reflected to some extent in the
large-scale statistics. Including the effects of fully hydrodynamical
in- and outflows was one of the main motivations for this work.

Another question raised by ZCTM10/11 is the redshift dependence of the
observed correlations. This is particularly important for HETDEX which
will cover a redshift range which is significantly below the one
explored by ZCTM10/11.

In this work, we revisit the the analysis of ZCTM10/11 using \lae RT on the
background of snapshots of the MareNostrum-Horizon simulation \citep{Ocvirk2008} at redshifts of
$z=2, 3$ and 4. The MareNostrum simulation has a
spatial resolution of 1 kpc (physical) and includes a model for
supernova feedback driving 
galactic outflows in a self-consistent fashion. In addition to
evaluating the correlations of the \lae observed fraction with the 
IGM density and velocity on linear scales, we tested for a possible
dependence on the orientation of the halos'
angular momentum relative to the line-of-sight, serving as a proxy for
the orientation of the galactic disk. We find a positive result,
indicating that tidal alignment of halos might give rise to additional
spurious signals in redshift space distortions \citep{Hirata09}. 

Our numerical techniques for \lae RT are
summarized in Sec. \ref{section_code} of the appendix.  We describe the 
details of the simulation and our
postprocessing runs in Sec. \ref{sec:sim}. Our results for
the correlations of large-scale density and velocity fields with \lae
observed fractions are presented in Sec.\ \ref{sec:results}.

\section{\lae Radiation Transport Calculations and Analysis}
\label{sec:sim}
\subsection{The Horizon-MareNostrum Galaxy Formation Simulation}
\label{sec:MNsim}

We applied our radiative transfer code LyS (see the appendix for details) 
to snapshots taken from the 
Horizon-MareNostrum Galaxy Formation run which was presented and described in
\cite{Ocvirk2008}.
The simulation was run using an updated version of the AMR code Ramses \citep{Teyssier2002}, 
including metal dependent cooling, star formation, a simple supernova feedback model and UV
heating. 
The box had a comoving size of 50 Mpc/h with a
physical resolution of 1 kpc. Star formation took place in the
interstellar medium (ISM), defined as gas with a number density greater than
0.1 $n_H$/cm$^3$. The dark matter particle mass was $8 \times 10^6$ M$_\odot$ with a total particle
count of $1024^3$. The simulation was run assuming a standard
  $\Lambda$CDM cosmology with $\Omega_M$ = 0.3, $\Omega_\Lambda$ =
  0.7, $\Omega_B$ = 0.045, $H_0$ = 70 km/s/Mpc and $\sigma_8$ =
  0.9. For more information on the spectroscopic properties of
  galaxies in the simulation, see \cite{Gay2010}.

\subsection{Preprocessing}
We rebuilt the AMR hierarchy of the
MareNostrum run
and calculated the temperatures of the gas cells from
the specific pressure assuming photoionization equilibrium
\citep{Katz1995}. In the ISM regions,
the breakdown of single component fluid description leads to
artificially high temperatures.
To overcome this problem, we 
enforce an upper limit of 2.5 $\times 10^4$ K on  the ISM temperature.

In order to find the emission spots for the Lyman-$\alpha$ photons, we used the
HOP algorithm \citep{Eisenstein1998} to produce a halo list. We used a
standard set of parameters ($\delta_{outer}=80, \delta_{saddle}=200, \delta_{peak}=240$) and rejected particles
groups that consist of less than 600 particles after the regrouping
process, corresponding to a cut-off mass of $4.8 \times 10^9
M_\odot$. 
With this cut-off, we have a sample size of $\sim$ 42.000/49.000/51.000 emitters at redshift 4/3/2. 
The mass range of these emitters is 5 $\times 10^9$ M$_\odot$ to 3.1 $\times 10^{12}$/8.0 $\times 10^{12}$/3.3 $\times 10^{13}$ M$_\odot$ for $z=4/3/2$. 

\section{Our Simulations and Analysis}
\label{sec:postprocess}

The \lae RT was run as a postprocessing step on simulation snapshots
at redshift $z = 2, 3$ and 
4. Additionally to the fiducial case, we re-ran our simulation at
redshift 4 with a) the Hubble flow, b) the peculiar velocity field and
c) both turned off for interpretation and comparison. To achieve this, we set the Hubble constant in eq. \ref{eq_hubble} to zero and/or set the total bulk velocity $v$ in eq. \ref{eq_lorentz} to zero so that the restframe of each gas cell is identical to the restframe of the emitter. The thermal motion of the gas is however not affected by this procedure.
We also ran the redshift 4 snapshot along three different lines of sight.
For details on the initialization of spectra and luminosities, we
refer the reader to Sec. \ref{section_code}. The spatial resolution of
our output array is 16.3 kpc/h (comoving), corresponding to
0.67/0.74/0.91'' 
at redshift 4/3/2.

The output matrix was converted into
physical fluxes and surface brightnesses. By integrating over the
spectral information of the output matrix, we obtained surface brightnesses
of each $(i_y,i_z)$-pixel. For each halo, we ran a friend-of-friend
algorithm to find the apparent luminosity of the source. If the pixel
covering the central position of the halo had a surface brightness
exceeding a threshold $\eta$, we added its flux to the flux of the source, and
connected adjacent pixels
that are above the threshold.

Using the total flux of each source obtained with this procedure, we defined the source's inferred apparent luminosity 

$L_{apparent}$.
As a result, we 
can compute the fraction  
\begin{equation}
\epsilon =  \frac{L_{apparent}}{L_{intrinsic}}
\end{equation}
of the intrinsic luminosity that was detected. Since this quantity
measures the part of the intrinsic luminosity that an observer would
see, it plays the role of an observed fraction, and hereafter we will
refer to it by this term. 
We caution the reader that in our case, the
difference between intrinsic and inferred luminosity is not due to
destruction of photons by dust, but due to the application of a
detection limit.  

To prevent source blending, we identified and ignored sources that would
swallow up other emitters during the post-processing, although these
blended source are quite rare ($\sim$ 5\% of the total number) and
did not affect our results very much.  

The chosen value of the surface brightness limit $\eta$ is somewhat arbitrary because we
didn't include dust and did not model the systematic errors of a real
observation in detail. We used a value similar as ZCTM10: 
\begin{equation}
 \eta =  5 \times 10^{-19} \textrm{erg s}^{-1} \textrm{cm}^{-2} \textrm{arcsec}^{-2}
\end{equation}
This particular value was chosen to be well above the noise level in
the output data. We caution the reader that this threshold is orders of magnitude smaller than the detection threshold of e.g. HETDEX, which is $\eta \simeq 10^{-17} \textrm{erg s}^{-1} \textrm{cm}^{-2} \textrm{arcsec}^{-2}$. We chose a lower value in order to be comparable with ZCTM10 but also not to degrade statistics by having only few sources detected.

The observed fraction of \lya along a specific line of sight is related to the density and velocity structure along the line of sight \citep{Wyithe2011a} (note that ZCTM10 and ZCTM11 stress the importance of the structure in the perpendicular directions). In order to find the correlations between the dark matter distribution in the
MareNostrum simulation and the observed fractions on linear scales relevant for LAE redshift surveys, we closely followed the strategy described by ZCTM10.
The dark matter particles were interpolated
onto a grid using a cloud-in-cell algorithm and smoothed 
out on a scale of 10/12/15 Mpc/$h$ with a top-hat filter of this diameter. We chose this filtering scale to obtain the density field in the linear regime at redshift 4/3/2 consistent with our calculations below. From the smoothed density
field, we calculated the linear velocity field and density/velocity
gradients along the line of sight.

The smoothed density and velocity fields are well described by linear
theory. From the continuity equation
\begin{equation}
 \dot \delta = - \frac{1}{a} div \; \vec u
\end{equation}\label{eq_continuity}
one finds the peculiar velocity field $\vec u_k$ 
in Fourier space:
\begin{equation}
 \vec u_k = fHa\frac{i \vec k }{k^2} \delta_k
\end{equation}
We are only interested in the line of sight component of the
velocity field which we assume to be parallel to the $x$-axis here:
\begin{equation}
 u_x = fHa\sum_{\vec k}\frac{k_x i}{k^2} \delta_k e^{i\vec k \cdot \vec r}\,\,.
\end{equation}
The spatial derivative of the velocity field in the line of sight 
is given by
\begin{equation}
 \frac{\partial u_x}{\partial x} = -fHa\sum_{\vec k} \frac{ k_x^2}{k^2} \delta_k e^{i\vec k \cdot \vec r}\,\,.
\end{equation}
We also calculate the angular momentum of the individual halos
directly from the particle data.

\section{Results}
\label{sec:results}

\subsection{Overview}
   \begin{figure}
   \centering
   \includegraphics[width=\linewidth]{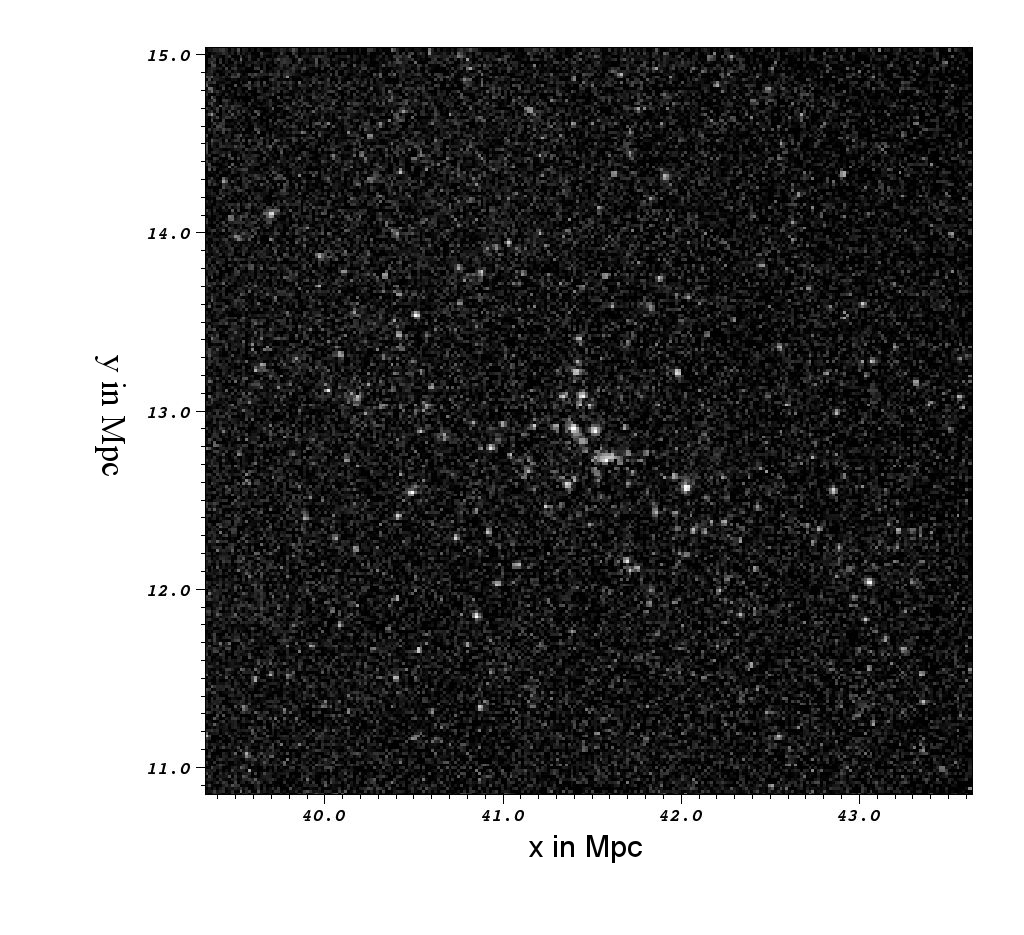}
      \caption{Lyman-$\alpha$ image of a small part ($\sim$5 $\times$
        5 Mpc) of the simulation volume at redshift 4. The observer
        is located along the positive $z$-axis, the snapshot
        corresponds to $z=4$.
              }
         \label{fig_overview_field}
   \end{figure}

   \begin{figure}
   \centering
   \includegraphics[width=\linewidth]{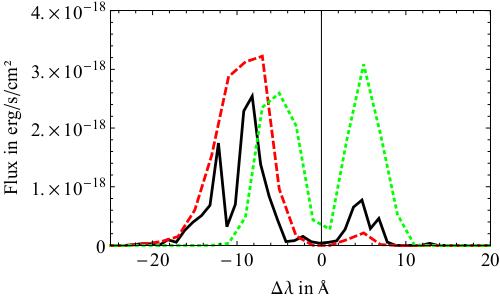}
      \caption{Typical Spectrum of an emitter with a mass of 1.9
        $\times$ 10$^{10}$ M$_\odot$. Shown is the spectrum for the
        fiducial simulation (solid line), the simulation without any
        peculiar velocities and Hubble flow (dotted line), and a
        simulation where peculiar velocities were enabled but the
        Hubble flow was switched off (dashed line). Wavelength is
        given with respect to the observer's restframe. For the fiducial simulation, the redshift due to the Hubble flow within the simulation box was ignored.
              }
         \label{fig_overview_spectrum}
   \end{figure}
Figure \ref{fig_overview_field} shows a spectrally integrated image
of the LAEs in the box as seen by an observer located along the
positive $z$-axis. In fig. \ref{fig_overview_spectrum}, the spatially
integrated spectrum of an emitter with a mass of $1.9 \times 10^{10}$
M$_\odot$ is shown for three different setups: the solid line shows
the spectrum of the fiducial run, the dotted line shows a run with all
peculiar motions and the Hubble flow artificially set to zero, and the dashed line is
obtained from a simulation where only the Hubble flow was switched off. 
For the case without peculiar motions and Hubble flow, we clearly see the
typical double-peaked spectrum that one would get from a static sphere
(see fig. \ref{fig_sphere}\footnote{Note that in
  fig. \ref{fig_overview_spectrum}, wavelength is shown instead of
  frequency.}). Deviations from this solution result from anisotropic density
fields. Turning on the velocity field, we obtain the typical spectrum
of an 
infalling sphere (see fig. \ref{fig_sphere_in}) which is intuitive
since the region in the halo's vicinity should show clear
infall. Photons are thereby shifted to the blue side of the spectrum,
undergoing only few scatterings after leaving the halo. If we switch
on the Hubble flow, the situation changes. Blue photons leaving the
ISM are shifted back into the line center and scattered in the
intervening IGM. As a consequence, the observed flux
is significantly reduced because photons
are scattered out of the line of sight. 
We note again that these photons are not 
destroyed by dust but they contribute to a noise level
of diffuse emission. The Hubble flow transports photons from the blue
to the red side of the spectrum. Once photons have left the line
center to the red side they will be further redshifted, 
making subsequent scatterings more and more improbable. 

\begin{figure}
   \centering
   \includegraphics[width=\linewidth]{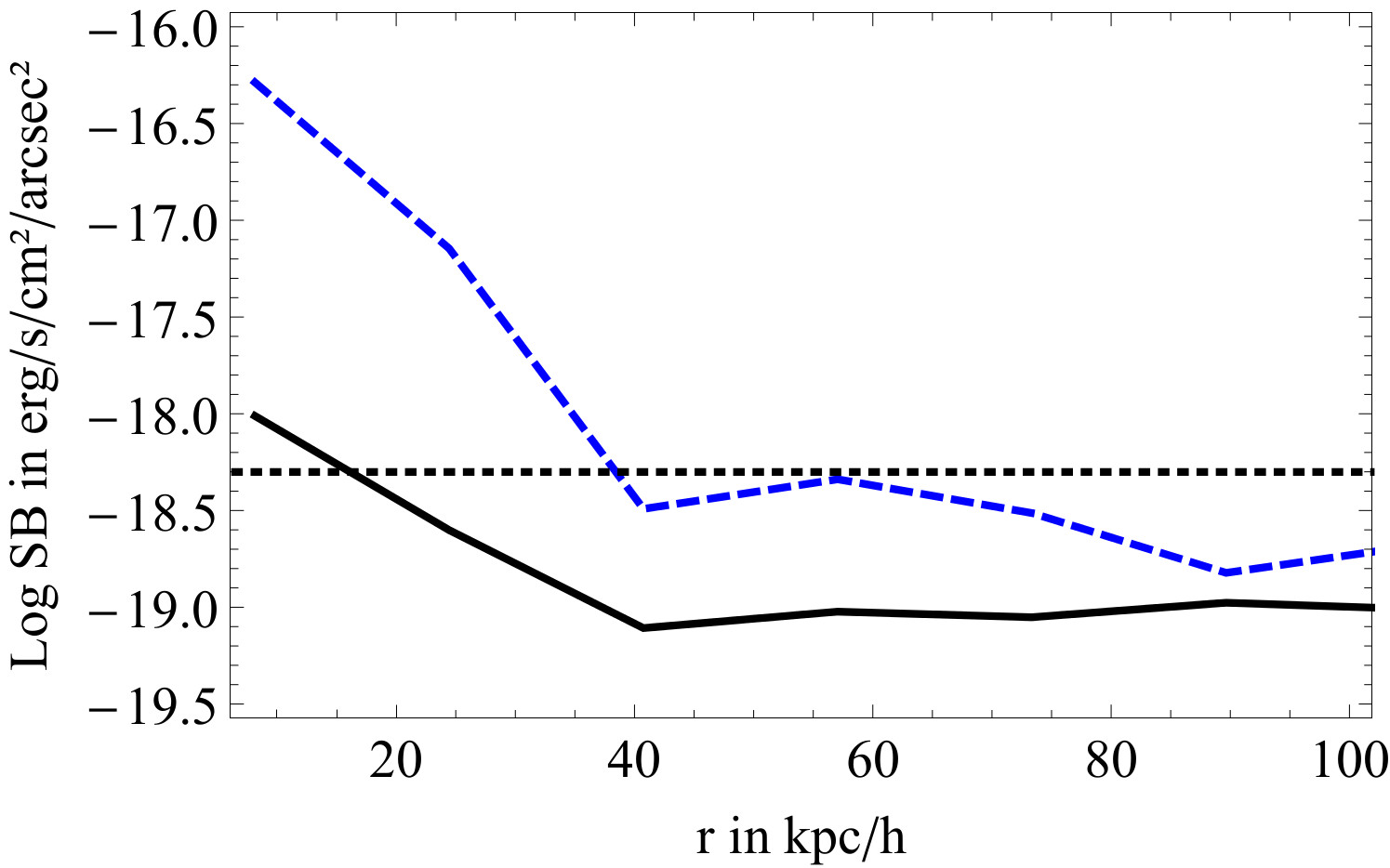}
      \caption{Surface brightness profiles for two sources with observed fraction of $\sim$ 30$\%$, with masses of 1.9 $\times$ $10^{10}$ $M_\odot$ (black solid line) and 7 $\times$ $10^{11}$ $M_\odot$ (blue dashed line). The horizontal dashed line indicates the detection threshold. On average, there were only 2 pixels per source above the detection threshold, and over 90\% are detected in less than 5.5 pixels.(Comment: This plot was added.) }
         \label{fig_sb}
   \end{figure}

The surface brightness profiles for two sources are shown in fig. \ref{fig_sb}. It is worth noting that while ZCTM10 reported extended \lae halos (r
$\sim$ 300 kpc) with surface brightnesses of $\sim$ 10$^{-20}$
 $\textrm{erg s}^{-1} \textrm{cm}^{-2} \textrm{arcsec}^{-2}$ at $z=5.7$ 
we find rather compact sources. These differences might be partly attributed to the lower redshift in our simulation. Additionally, as we resolve the ISM at least marginally, most of the scatterings happen in the ISM where the optical depth is high due to high HI densities. This shifts the photons out of resonance and reduces the optical depth of the immediate surroundings of the halo. Since later IGM scatterings happen far away from the emitting halo, they contribute to a diffuse background rather than to an extended \lya halo. 

     \begin{figure}
   \centering
   \includegraphics[width=\linewidth]{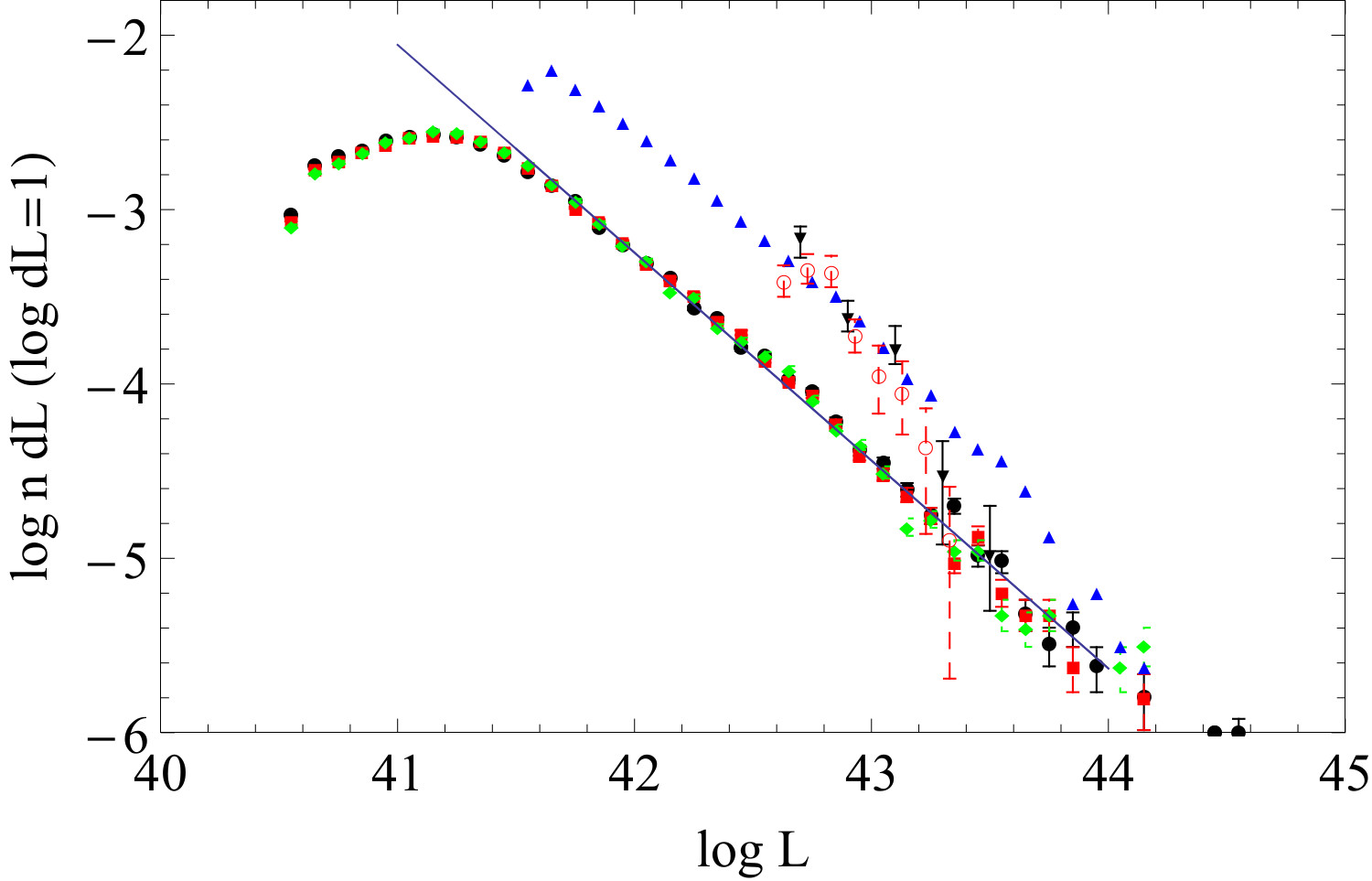}
      \caption{Luminosity function from our simulation at $z=4$ for
        three different lines of sight. The observer is set along the
        x-axis (circles), y-axis (boxes) and z-axis (diamonds), 
        respectively. Triangles without error bars show the
        distribution of the intrinsic luminosity. Hollow circles show the observed luminosity function for $z=4.5$ \citep{Wang2009}, the flipped triangles show the luminosity function for $z=3.7$ as observed by \cite{Ouchi2008}. The line depicts the
        log-linear fit to the data, yielding a power law with
        $\alpha=-1.19$. 
      Number density is in units of Mpc$^{-3}/\log(L)$. 
              }
         \label{fig_lfunction}
   \end{figure}
   
One can also obtain the luminosity function for our simulation, shown in fig. \ref{fig_lfunction} for the three different lines of sight at redshift 4. We also show the intrinsic luminosity function (triangles) of our simulation. The overall shape of the luminosity function is not changed by the RT process. The drop at the low luminosity is due to incompleteness, since LAEs that are below the detection limit have an apparent luminosity of zero.
The detection threshold and the
assigned intrinsic luminosity introduce a free parameter in our
model, shifting the luminosity function by a constant factor. Since we
are mostly interested in changes of the observed fraction relative to
the mean, we ignore this shift here.

\subsection{Correlations between Large Scale Structure and Observed Fraction}
In this section, we focus on our results from the redshift 4 snapshot. 

\begin{figure*}
   \centering
   \includegraphics[width=7cm]{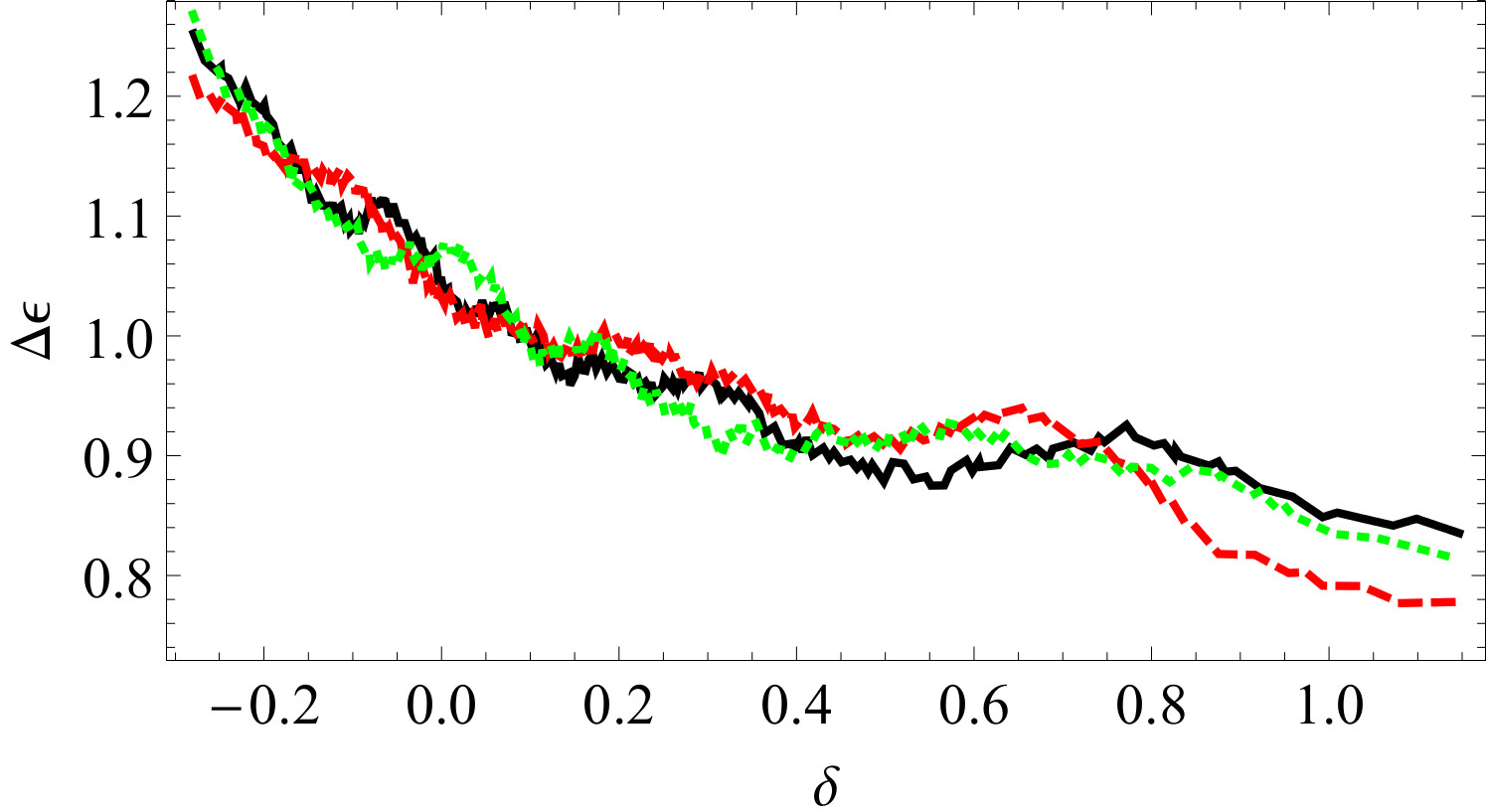}
\includegraphics[width=7cm]{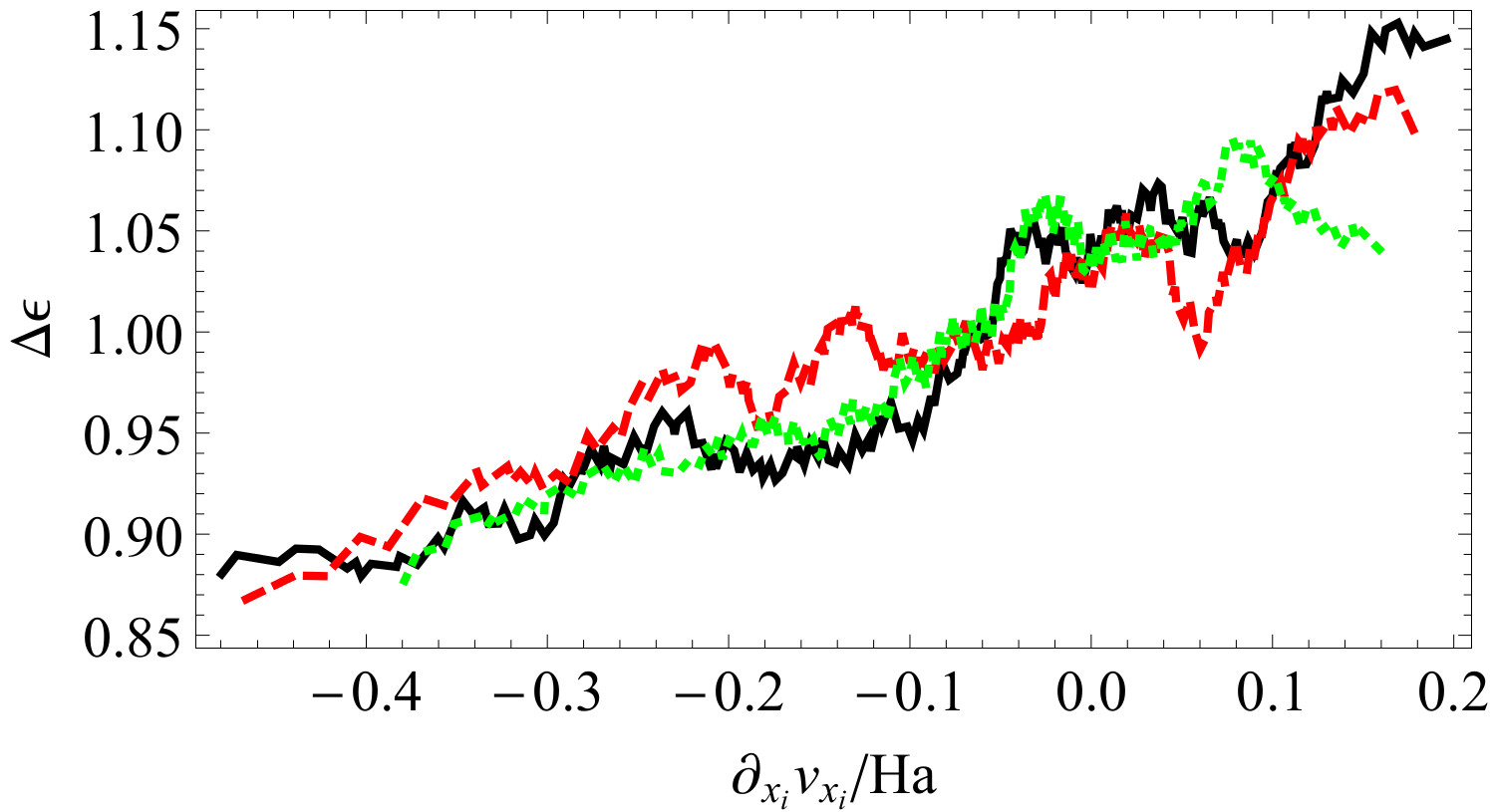}
\vspace{0.4cm}

\includegraphics[width=7cm]{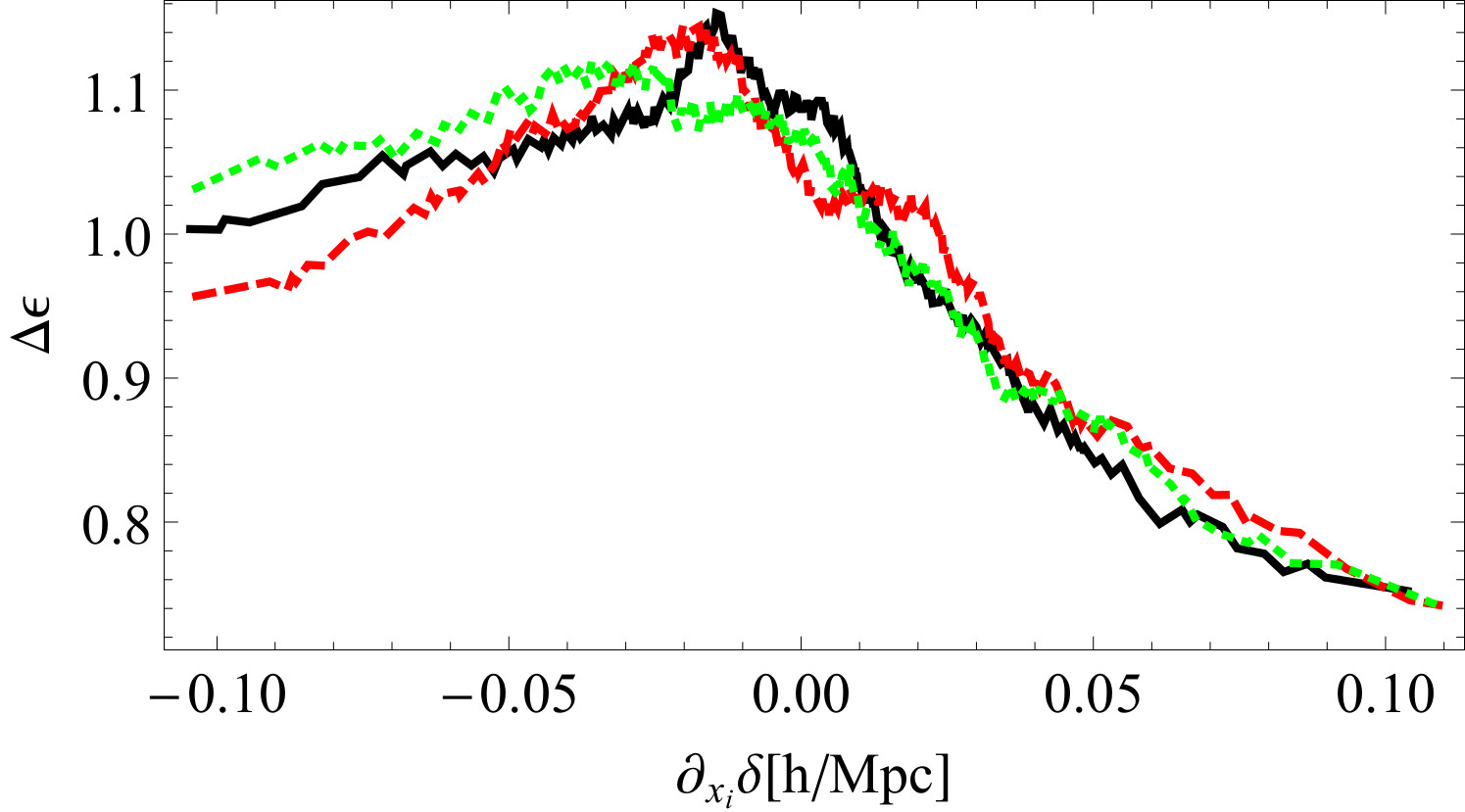}
   \includegraphics[width=7cm]{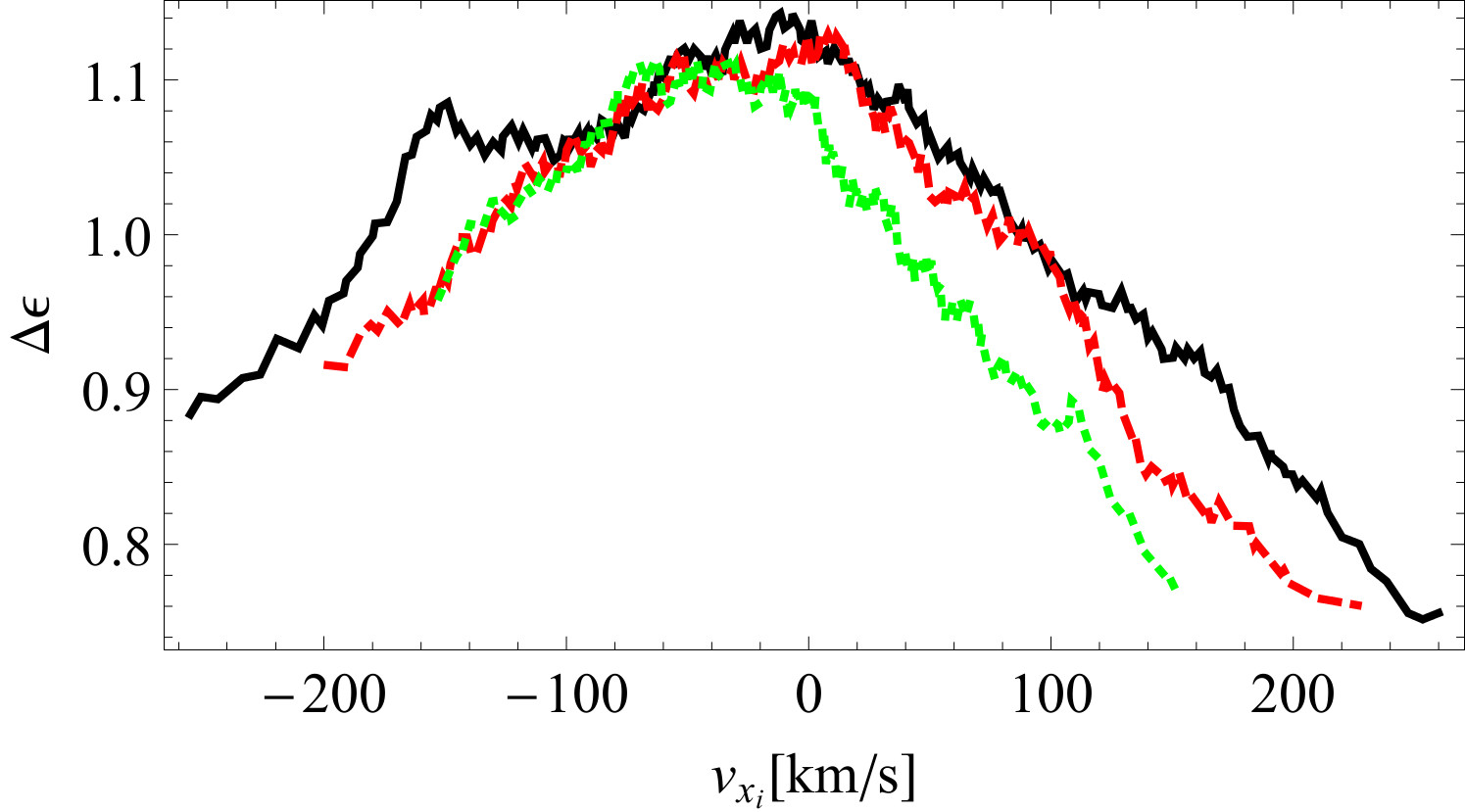}
      \caption{
        Correlations between observed
        fraction and large scale structure. On the $y$-axis, the
        relative observed fraction is shown in all four plots. The
        $x$-axis shows density (upper left), the line of sight velocity gradient (upper right), the line of sight 
        density gradient (bottom
        left) and the line of sight velocity (bottom right). The three different
        lines correspond to the 
        observer being located along the $x$-axis (black, solid), $y$-axis
        (red, dashed) and $z$-axis (green, dotted).
        } 
         \label{fig_3lines}
   \end{figure*}

Figure \ref{fig_3lines} shows how the observed fractions correlate
with the dark matter overdensity $\delta$, the density gradient along
the line of sight, the line of sight velocity and the line of sight
velocity gradient, all evaluated at the positions of the sources for three different lines of sight. The observed fraction is given
relative to the mean observed fraction, 
\begin{equation}
 \Delta \epsilon = \frac{\epsilon}{\bar \epsilon}\,\,.
\end{equation}

The mean observed fraction $\bar \epsilon$ is 30/63/87\% for our
fiducial runs at redshift 4/3/2. ZCTM10 find a much lower mean observed
fraction of a few percent at redshift 5.7. \cite{Laursen2011} also calculated observed fractions from nine simulated galaxies at 3.5. Although their sample is small, their mean observed fraction is around 24\% (also note they do include photon destruction by dust). 

We use the full sample of emitters and generate the plots of the correlations applying a moving average to the data set, averaging over 4000 emitters per data point. As described above, densities and
(linear) velocities are obtained from the smoothed dark matter
particle data. 
Density gradient, velocity and velocity gradient
plots depend on the line of sight chosen, so we plot the relevant
component $x_i$ or the derivative with respect to $x_i$, where $x_i = x,y,z$ 
for observers located along the respective axis.  

The correlation between density and velocity
gradient follows directly from the continuity equation (see
eq. \ref{eq_continuity}). Statistically, this also holds for the
individual components of the divergence, i.e., the line of sight
velocity gradient. The correlation between line of sight velocity and
line of sight density gradient is quite intuitive: halos beyond a
large scale overdensity move towards it and hence towards the
observer, and vice versa. Both correlations are shown in fig. \ref{fig_cross_vz_delta_dz} and fig. \ref{fig_cross_delta_vz_dz}. Because of this direct connection between the
two pairs of observables, we discuss each of the pairs together. 

\begin{figure}
   \centering
   \includegraphics[width=\linewidth]{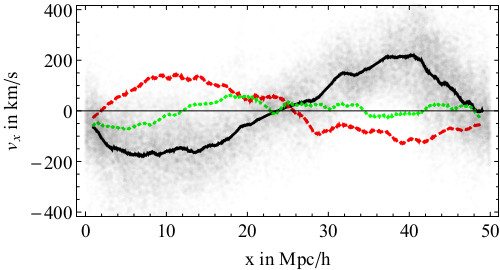}
      \caption{Averaged line of sight component of the halo velocities versus their location along the line of sight at  $z=4$ for three lines of sight parallel to the $x$-/$y$-/$z$-axis (black solid/red dashed/green dotted).To illustrate the scatter, the distribution of halos is plotted in gray for the $x$-axis data.
              }
         \label{fig_distribution}

\vspace{0.2cm}
   \centering
   \includegraphics[width=\linewidth]{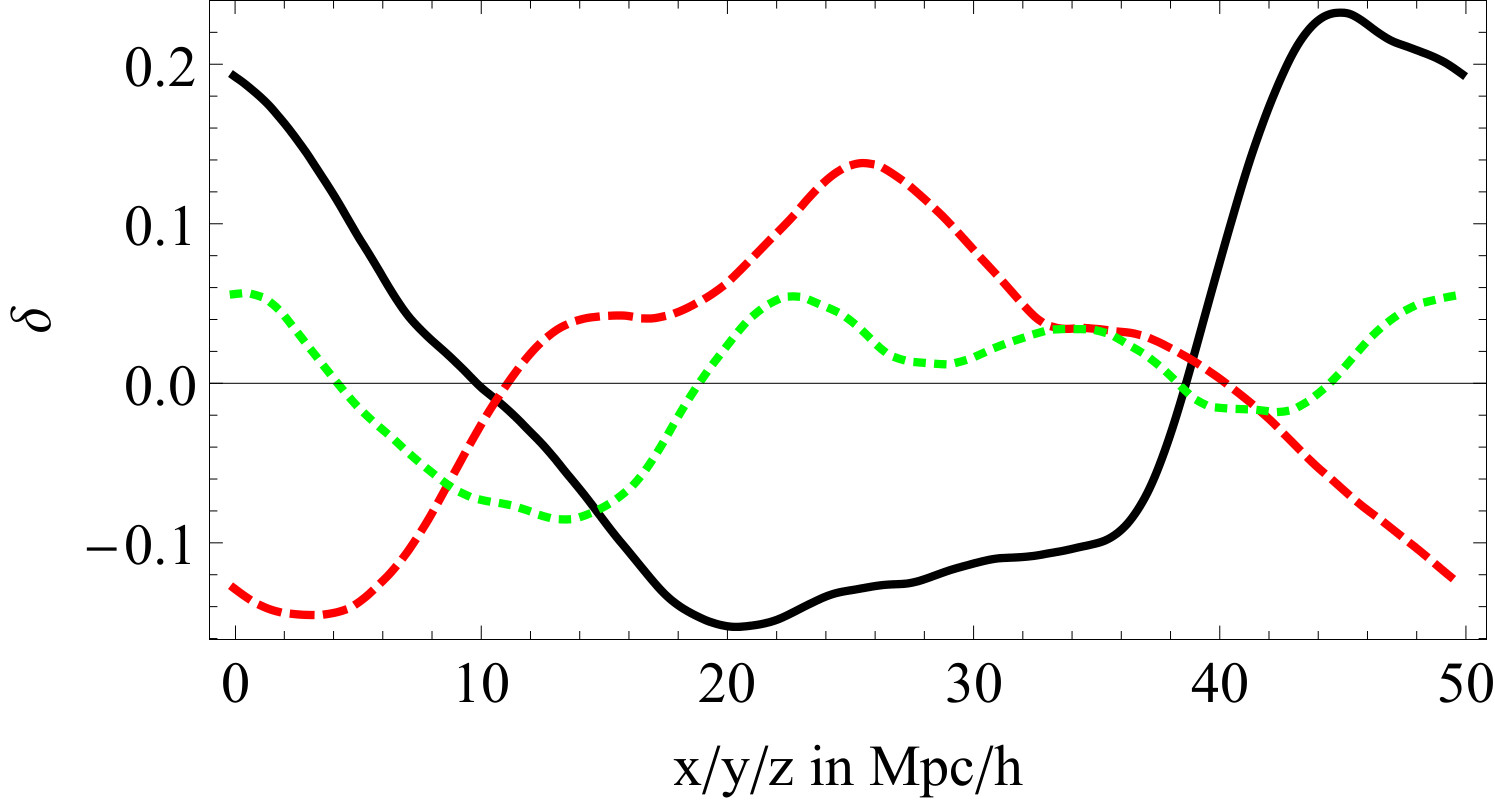}
      \caption{Projected dark matter density along the $x$/$y$/$z$-axis (black solid, red dotted, green dashed line), smoothed on a scale of 10 Mpc/h for $z=4$. 
              }
         \label{fig_deltadistribution}
   \end{figure}

As can be seen in fig. \ref{fig_3lines}, correlations differ between
different lines of sight. Although the trends are mutually consistent,
deviations of up to 10\% are clearly visible. We interpret this as
a consequence of cosmic variance. Investigation indeed shows
that the halo distribution and velocity fields clearly differ among
different lines of sight, which can be expected for a box of this
size. As an example, in fig. \ref{fig_distribution}, the $x/y/z$-component of the
halos' velocity is plotted against the $x/y/z$-coordinate. For this plot,
velocities are directly obtained from the dark matter particles, but
the linear approximation we use to build the correlation plots shows
the same behavior, i.e., a large scale, sine-like signal. This velocity distribution reflects the density structure in the box, as can be seen by comparison with fig. \ref{fig_deltadistribution}.

  \begin{figure*}
   \centering
   \includegraphics[width=7cm]{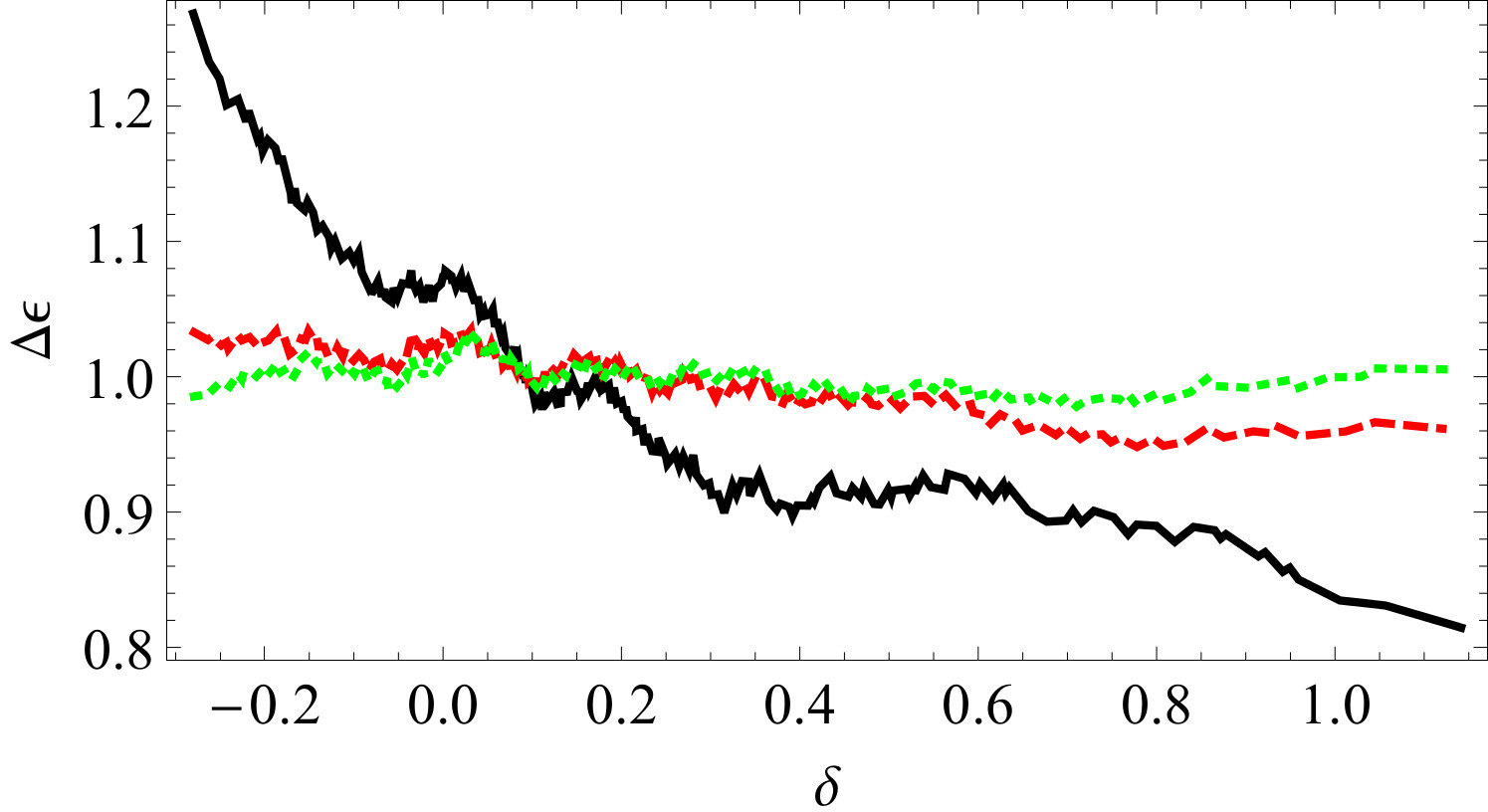}
\includegraphics[width=7cm]{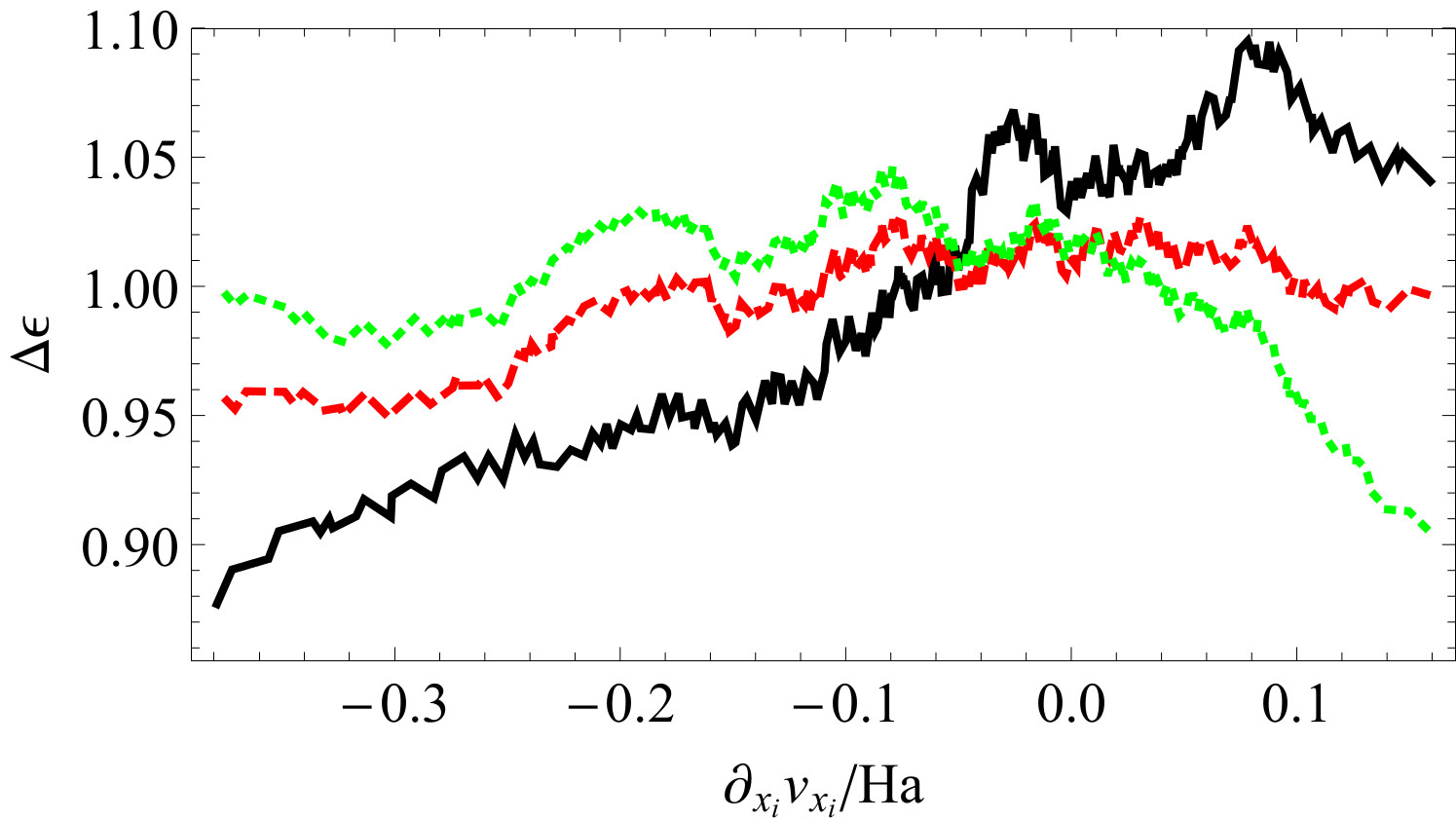}
\vspace{0.4cm}

\includegraphics[width=7cm]{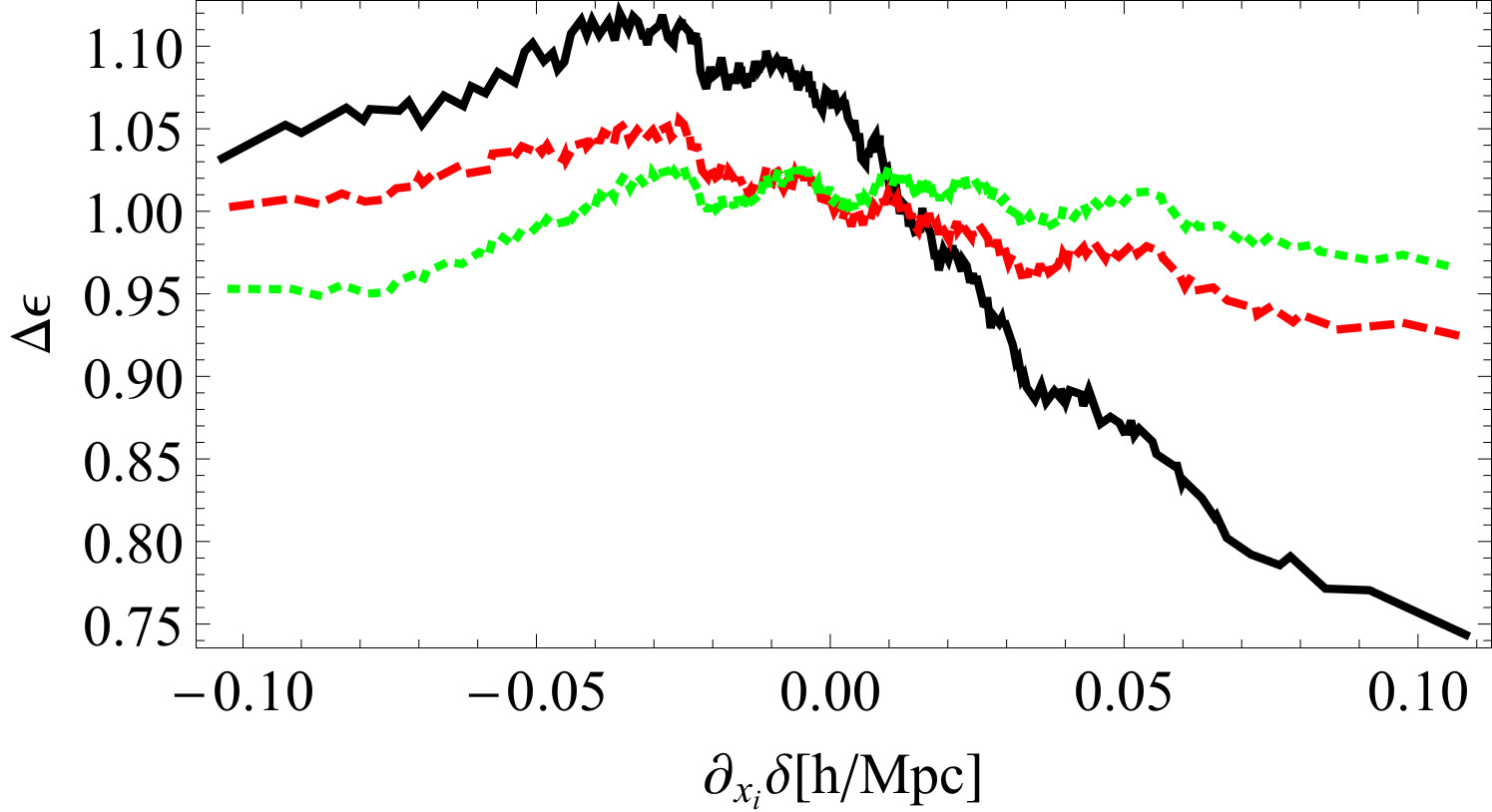}
   \includegraphics[width=7cm]{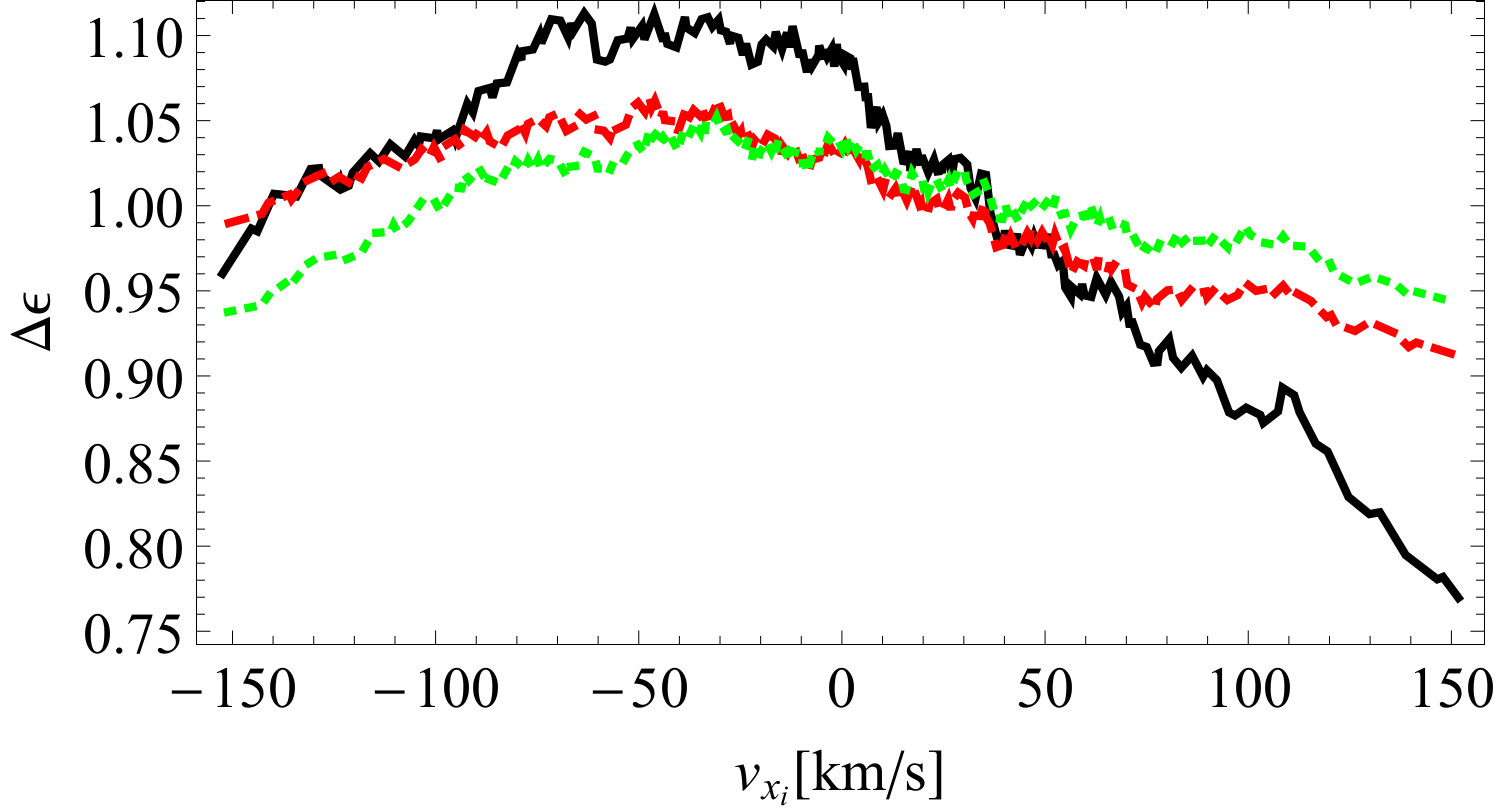}

      \caption{Same as fig. \ref{fig_3lines}, but here the lines
        correspond to the fiducial case (black, solid), a run without
        peculiar motions and Hubble flow (red, dashed) and a
        simulation with peculiar velocities, but without Hubble flow
        (green, dotted), as seen by an observer along the $z$-axis.} 
         \label{fig_compare}
   \end{figure*}

  \begin{figure*}
   \centering
   \includegraphics[width=7cm]{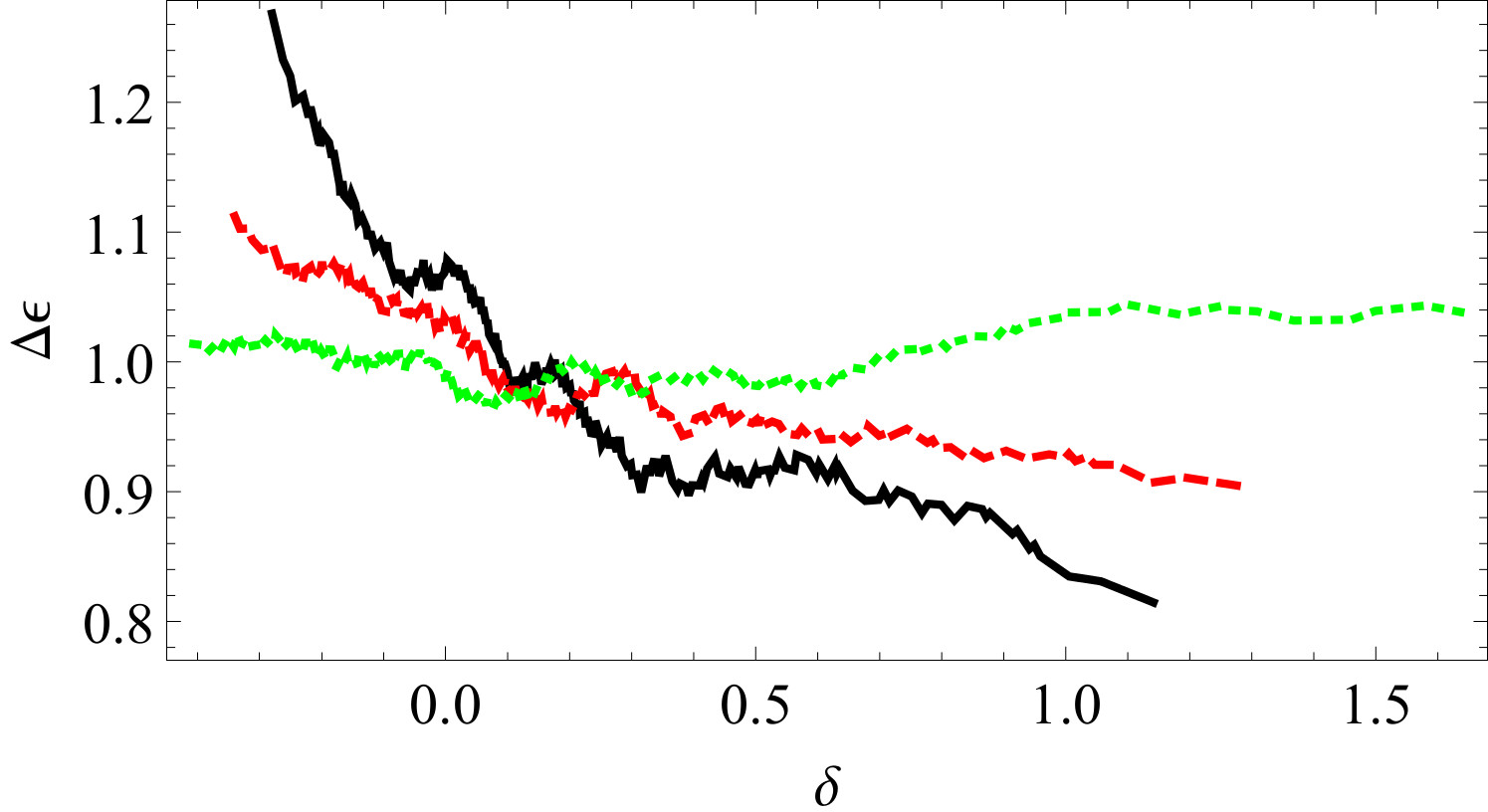}
\includegraphics[width=7cm]{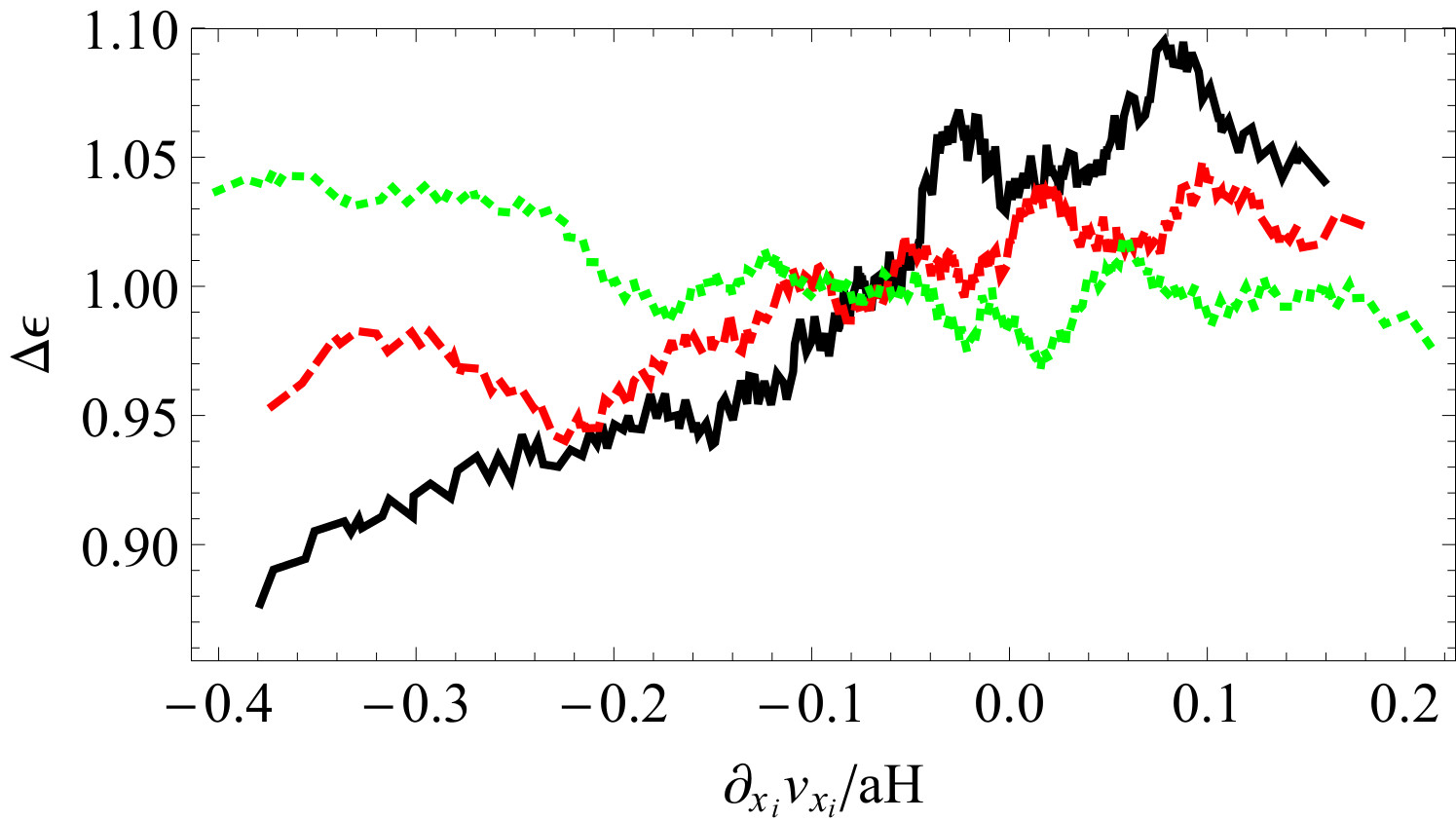}
\vspace{0.4cm}

\includegraphics[width=7cm]{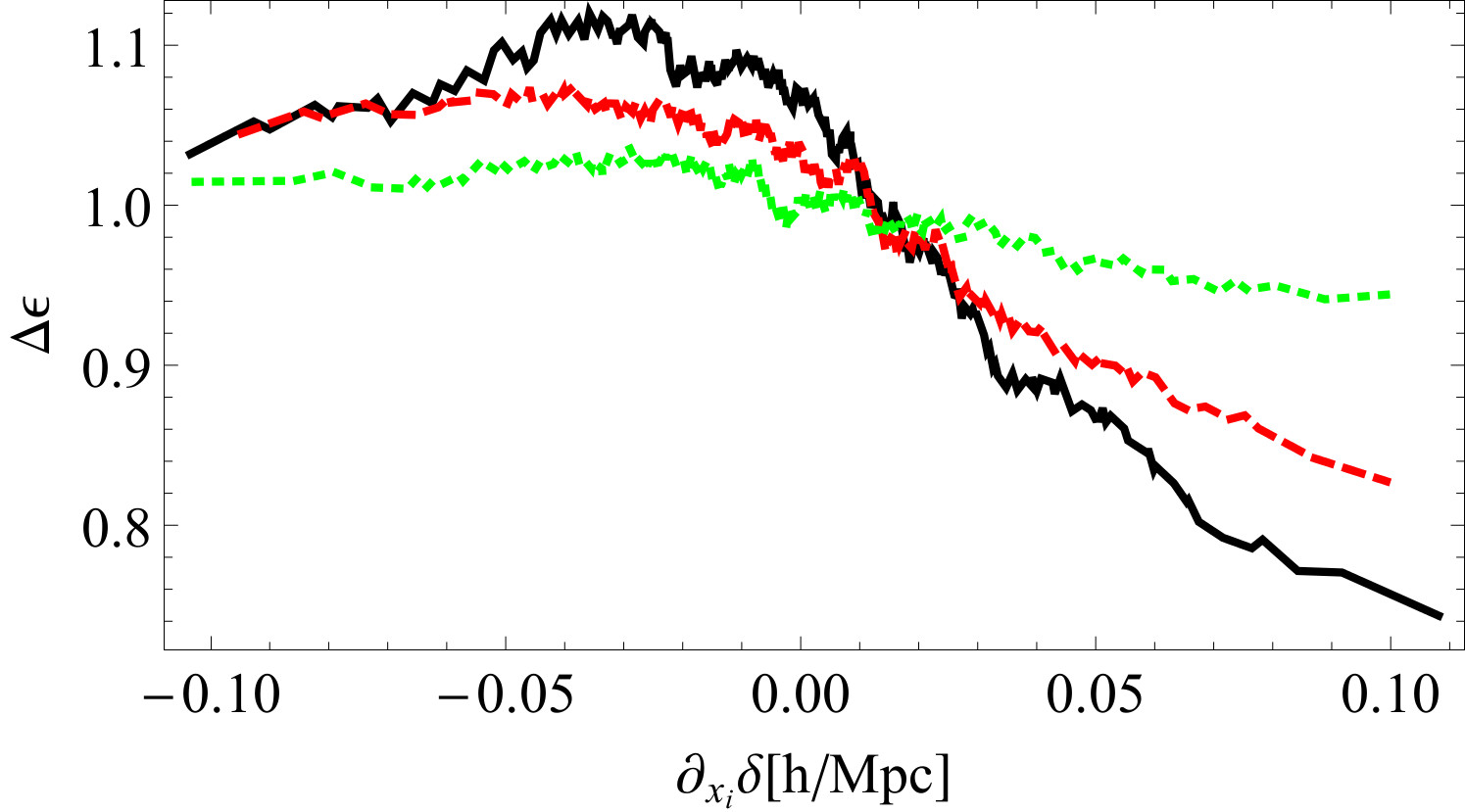}
\includegraphics[width=7cm]{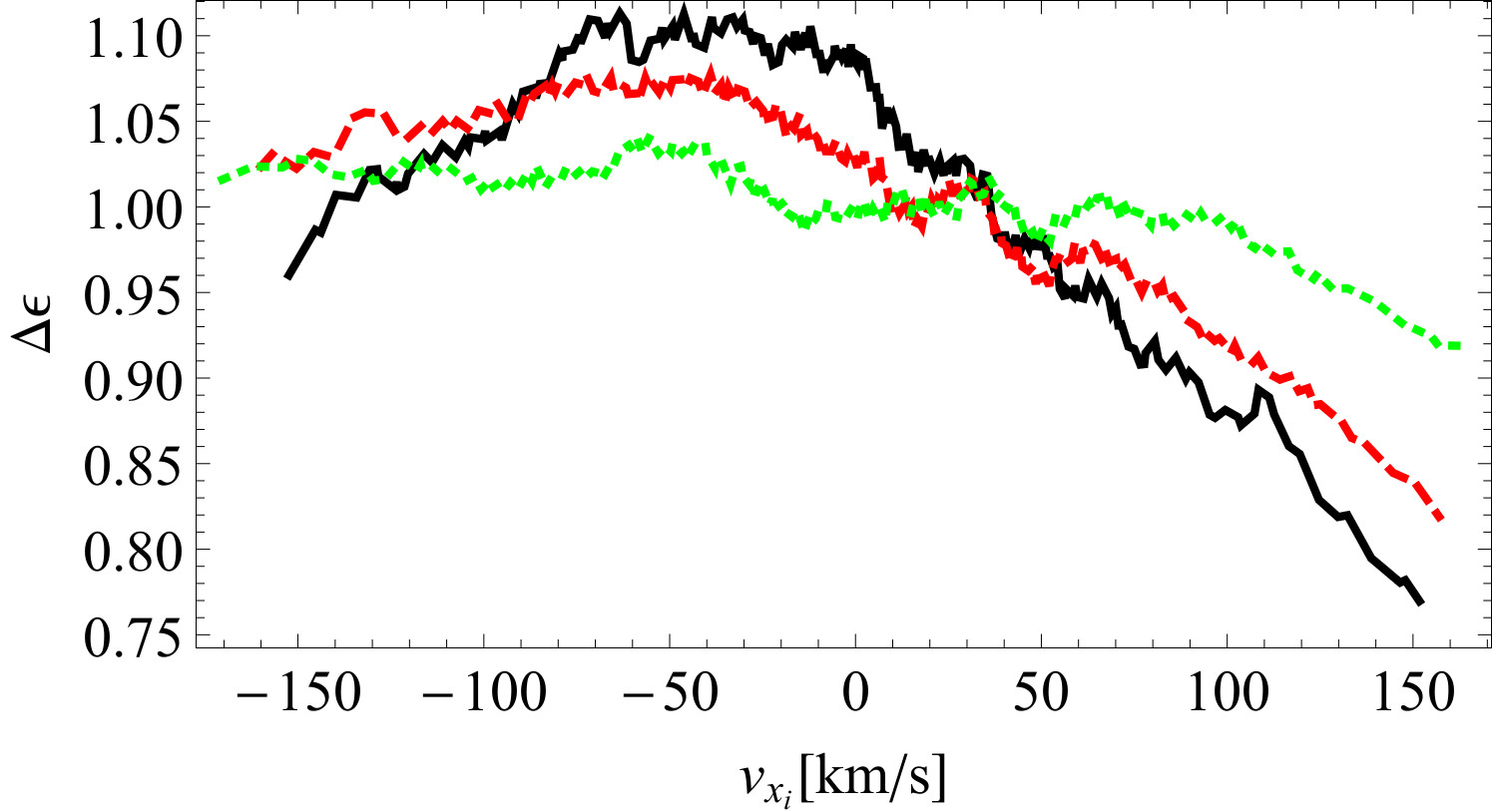}
      \caption{Same as fig. \ref{fig_3lines}, showing the redshift
        dependence of the correlations. The data for a redshift of $z=4$ (black, solid), $z=3$ (red, dashed)
        and $z=2$ (green, dotted) is shown. The observer is located
        along the $z$-axis.} 
         \label{fig_redshift}
   \end{figure*}

\subsubsection{Density}
As can be seen in the top left plot in fig. \ref{fig_3lines}, larger
overdensities are correlated with lower observed fractions. The effect
is quite strong with about 30\% in amplitude. We interpret this as a
result of diffuse scattering around halos in overdense
regions. Interestingly, the signal is highly suppressed if we switch
off either the Hubble flow or the peculiar velocity field (see fig.
\ref{fig_compare}, upper left subplot). Turning off the velocity field
renders the spectra nearly symmetric. Photons that leave the ISM are
not subject to further scattering, because 
neither local gas flows nor Hubble expansion can shift the photons
back into the line center. 
This interpretation is also supported by
calculations of the mean optical depth of the box. While in the line
center, the optical depth (assuming a temperature of $2 \times 10^4$ K and
taking the mean HI density from the simulation volume) is still
$\tau\sim 10^2$, a shift of 1 $\AA{}$ (in the local frame at $z=4$)
reduces the optical depth to $\tau \sim 10^{-3}$. This also makes
clear why velocity fields are so crucial in the radiative transfer.  

When we turn on the peculiar velocity, the signal remains weak, but
there is a slight decrease in observed fraction for halos in
underdense regions. We interpret this as a result of the fact that in
those underdense regions, small halos are dominant. The ISM in small
halos doesn't push the photons as far out into the wings as in the
larger ones, so the probability of being scattered in the IGM is
increased. Switching on the Hubble flow as well leads to our fiducial
case: the Hubble flow together with the peculiar velocities leads to
more numerous scatterings in overdense regions 
compared to the underdense regions.

\subsubsection{Line of sight velocity gradient}
\begin{figure}
   \centering
   \includegraphics[width=\linewidth]{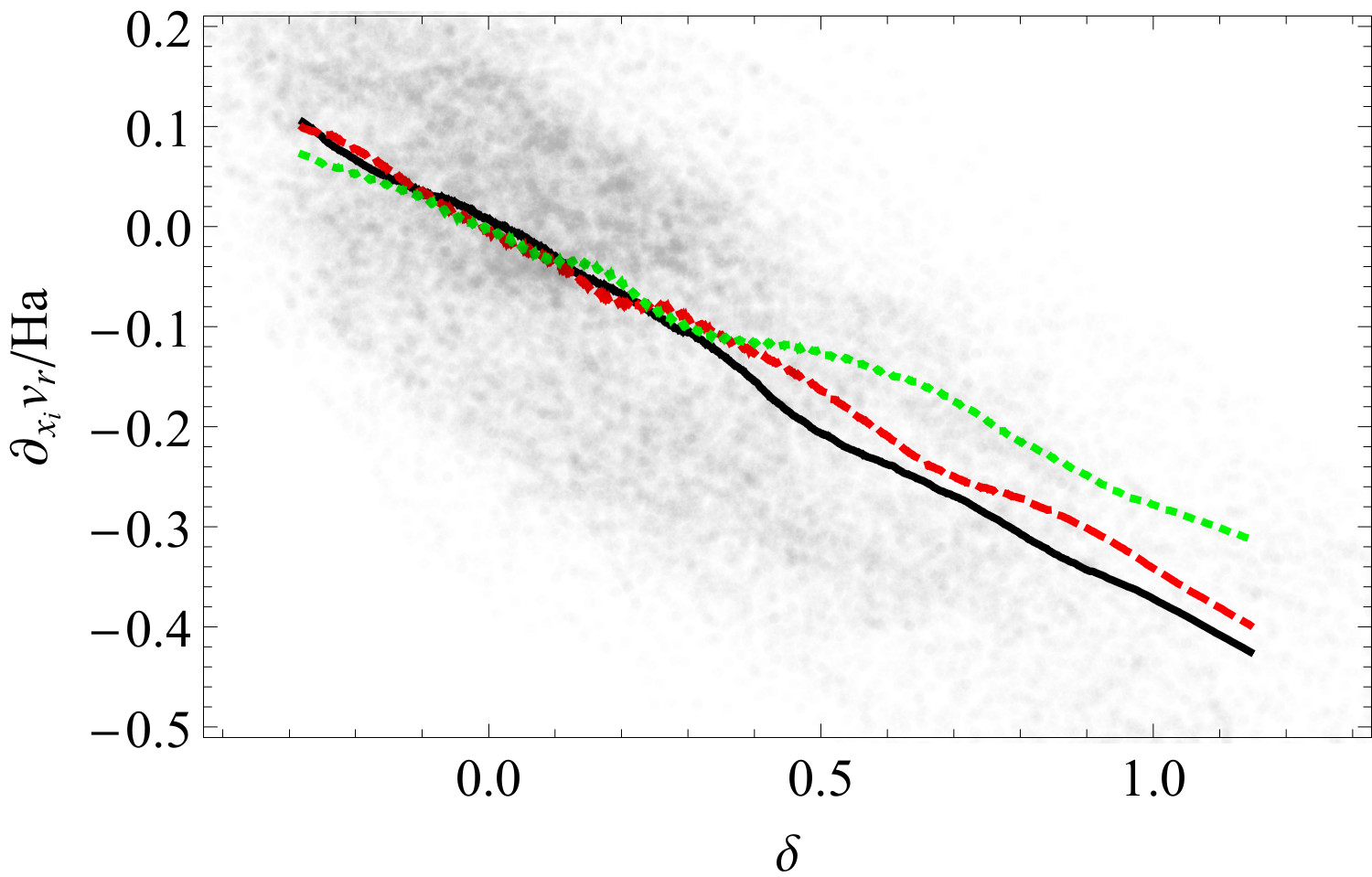}
      \caption{Correlation between the density and the line of sight velocity gradient for all three lines of sight at $z=4$, x-axis (black, solid), y-axis (red, dashed) and z-axis (green, dotted). The gray dots show the scatter, each point representing one halo (for the x-axis data). See text for details. }
         \label{fig_cross_delta_vz_dz}
   \end{figure}

The top right plot shows the correlation between line of sight
velocity gradients and observed fractions.
Here, we obtain
qualitatively the same result as ZCTM10, although with a much lower
amplitude; larger velocity gradients lead to a higher observed
fraction. 
As has been discussed above, one can see that 
the effects of
density and velocity gradient are anti-correlated 
as expected from linear theory (cf.\ fig.\
\ref{fig_cross_delta_vz_dz}). For this reason, there is no meaningful
way to completely disentangle the two physical mechanisms. 
However, the dominant effect from the velocity gradient
can be understood from the fact that a
higher velocity gradient corresponds to a higher local effective
Hubble flow. Since most of the photons that leave a halo are blue, the
higher the effective Hubble flow, 
the faster the photons are transported
through the line center and into the red wing. If we turn off the
Hubble flow, the effect is reversed: Now, a positive local gradient
suppresses the emitters. This might be due to the fact that in this
case, the redshifting through the line center takes place on a much
larger length scale because the peculiar velocity gradient is much
smaller than the Hubble flow. 

It is worth  pointing out that while in our simulations the amplitude
of the correlation with line of sight velocity gradient is of the same
order of magnitude as for the other correlations examined, ZCTM10 find a
larger  amplitude for the line of sight velocity gradient by about one order
of magnitude. There are several factors that might be at play
here. First of all, ZCTM10 worked at a higher redshift of $z = 5.7$. The
mean density 
is about 2.5 times higher at that redshift and therefore,
the optical depth in the IGM is naturally higher, leading to more
scatterings in the diffuse large scale environment. This might leave a
stronger imprint of the bulk velocity fields (and their spatial
evolution) in the observed fraction. Secondly, 
whereas ZCTM10 semi-analytically map the baryons
onto the results of a pure $N$-body simulation, our gas distribution
follows from a high-resolution hydrodynamical simulation that includes
the effects of nonlinear in- and outflows which reduce the
correlations with flows on linear scales.
Additionally, we resolve the ISM at least
marginally, leading to a large number of scatterings in the dense
regions where the photons are emitted. After being processed through
the ISM, the photons have already been shifted from the line center to some extent. This also leads to
less scattering in the IGM, further reducing the impact
of the large-scale velocity field on the
observed fraction. 

\subsubsection{Line of sight density gradient}

This correlation is shown in the 
bottom left plot of fig. \ref{fig_3lines}. Due to the orientation of
the observer towards 
the box, negative values here indicate that the density decreases
in the direction of the observer. This makes the general trend in the
plot plausible: observed fractions are lower if there is an
intervening large scale overdensity region between the emitter and the
observer. The effect becomes smaller when turning off Hubble flow and
peculiar velocities, as can be seen in the lower left plot of
fig. \ref{fig_compare}. This is probably due to the fact that the
density of the environment becomes less important, as discussed in the
preceding paragraphs. In contrast to ZCTM10, we find a
  prominent peak in the density gradient signal, at a value of around
  zero. 
We attribute this effect to the fact 
that halos with a density gradient of $\sim$ 0 are 
predominantly located inside lower density regions and voids.
Due to the correlation of higher observed fractions with lower densities, those halos have a higher observed fraction. Analysis indeed shows that those low density halos ($\delta<0$) populate the region around $\partial_{x_i}\delta$ $\sim$ 0, and that their mean observed fraction is about 6\% higher than the mean observed fraction of halos in denser regions. The mean density at the location of halos with a density gradient between -0.03 and 0.0 h/Mpc, for example, is reduced by a factor of 50\% with respect to the full sample. Additionally, one can also notice that large absolute density gradients correspond to halos that are near to the large scale overdensity, and therefore in a region where the density is increased. For example, the mean large scale overdensity for halos with a density gradient $<-0.08$ h/Mpc is roughly twice the mean of the whole sample. Since this increases the optical depth also for the halos on the near side of the halo, it could also account for the dip on at large 
negative density gradients. This hypothesis is also supported by the fact that the effect does not fully vanish when the velocity field is set to zero, cf. fig. \ref{fig_compare}.
\subsubsection{Line of sight velocity}
\begin{figure}
   \centering
   \includegraphics[width=\linewidth]{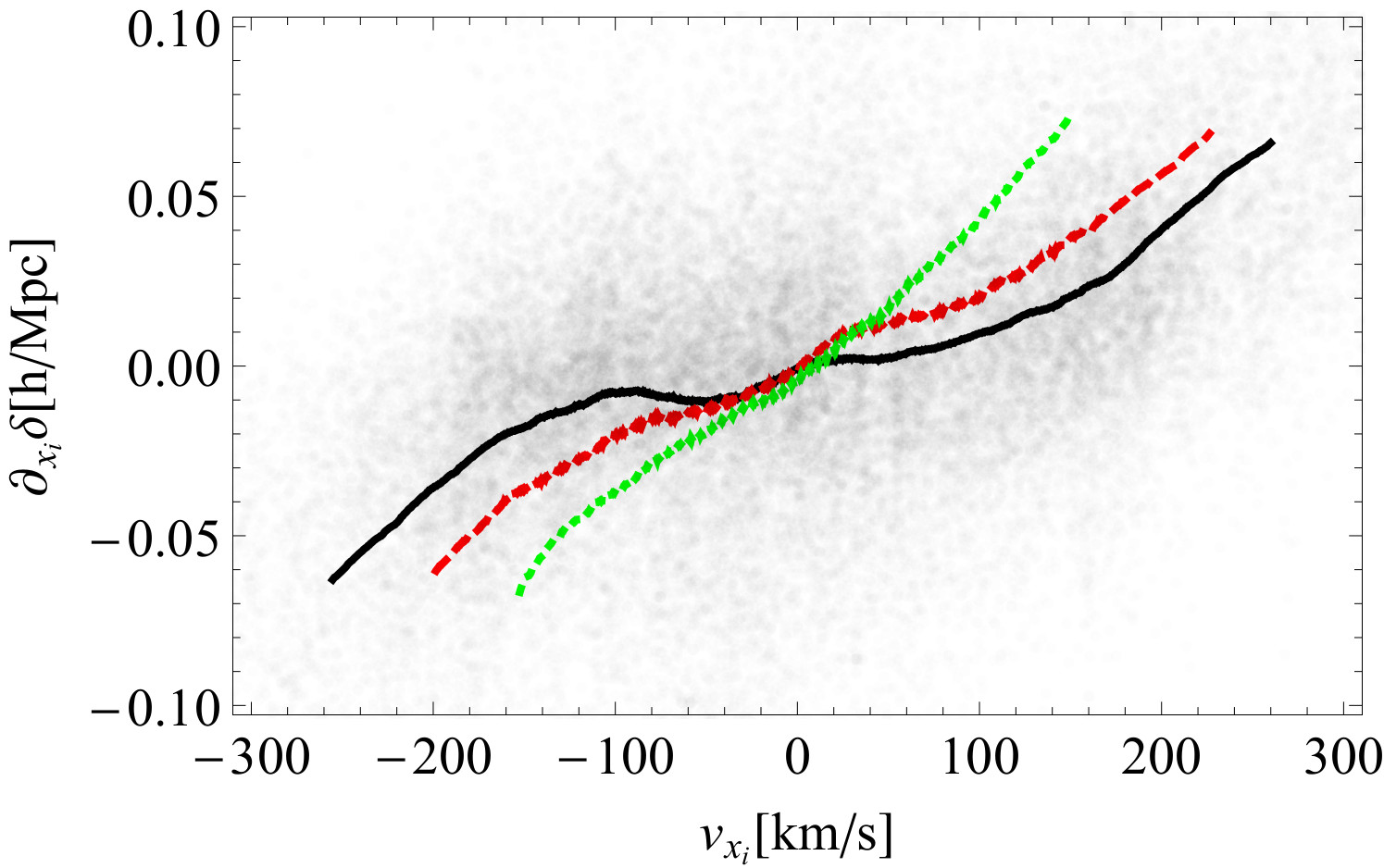}
      \caption{Here, we show the correlation between the line of sight density gradient and the line of sight velocity for all three lines of sight at $z=4$, x-axis (black, solid), y-axis (red, dashed) and z-axis (green, dotted). The gray dots show the scatter, each point representing one halo (for the x-axis data). See text for details. }
         \label{fig_cross_vz_delta_dz}
   \end{figure}

In the bottom
right plot, the correlation between line of sight
velocities and observed fractions is shown. The box is orientated so
that halos moving into the direction of the observer have positive
velocities. The plot's shape and the peak at around zero stays the
same even if we turn off the peculiar velocities in the simulation, as
can be seen in the lower right plot in fig. \ref{fig_compare}. This
indicates that the correlation is dominated by 
the density gradient which looks nearly the same as the
velocity signal when we turn off peculiar velocities. This is quite
plausible since density gradient and velocity fields 
are highly correlated, cf.\ fig.\ \ref{fig_cross_vz_delta_dz}.
With Hubble flow and peculiar velocities switched on, we see a much larger amplitude
and a strong suppression at positive velocities. Since halos with
positive velocity are moving towards the observer while those with
negative velocities are receding, we interpret this as a
consequence of the Hubble flow that suppresses blue halos more
strongly than the red ones. 


\subsection{Evolution with Redshift}
\label{sec:redshifts}
In fig. \ref{fig_redshift}, the results for the dark matter
correlation are shown for the redshift $z = 2, 3, 4$. For this part of
the analysis, the smoothing scale of the dark matter particles was
adjusted to stay in the linear regime at lower redshifts. As has been
discussed in a 
preceding section, the detection limit and the prescription of the
intrinsic luminosity are somewhat arbitrary. Since the total mean
density of the universe increases with redshift, we get different mean
observed fractions at lower redshifts. At redshift 4, we have a total
observed fraction of 30\%, at redshift 2 it has risen to above 80\%. For
individual emitters, the observed fraction cannot be much larger than unity,
so a higher total observed fraction can result in a compression of the
correlation signal.

 We find that correlations  decline in amplitude with
decreasing
 redshift. Since the mean density of the IGM decreases, the
 influence of the environment of the halos weakens.
For the correlation of the observed fraction with the line of sight
 velocity, the drop of the Hubble rate
 from $\sim$ 400 km/s/Mpc at $z=4$ to $\sim$ 200 km/s/Mpc at $z=2$ 
further reduces 
the influence of the Hubble flow on the observed fractions
 from halos that move towards the observer 
($v_r > 0$). 

 While for the density gradient and 
the line of sight velocity, this decline in amplitude preserves the
 overall trend, for the density and the line of sight velocity
 gradient the lowest redshift $z = 2$ shows a slight turnaround. For
 this redshift, halos in dense regions and in regions
 with a smaller velocity gradient are preferred by a few
 percent. One possible explanation is that a this later stage of structure formation, the
 medium in those dense regions is hotter and therefore contains less
 neutral gas. Since overdensities and low velocity gradients are
 coupled, this also affects the velocity gradient correlation. 

In table \ref{tab1}, we show the results for linear fits to the
general trends seen in the correlations. 
They were calculated ignoring the dip on the left side in the density gradient and velocity plots (i.e. the
left edge of the velocity- and density gradient range). Due to the
nonlinearity of the density correlation, the fitted value for this
plot strongly depends on the chosen range for the fit. We obtained our
fitted value by ignoring the steep decline below $\delta = 0$.  

\begin{table}
\caption{Evolution with redshift}             % title of Table
\label{table:1}      % is used to refer this table in the text
\centering                          % used for centering table
\begin{tabular}{c c c c}        % centered columns (4 columns)
\hline\hline                 % inserts double horizontal lines
 & z=4 & z=3 & z=2 \\    % table heading 
\hline                        % inserts single horizontal line
   $\partial_\delta \Delta \epsilon$ & -0.48 & -0.15 & -0.05 \\      % inserting body of the table
   $\partial_{\partial_r \delta} \Delta \epsilon$ & -3.0 & -1.94 & -0.78 \\
   $\partial_{v_r} \Delta \epsilon$ & 1.9 $\times 10^{-3}$&  1.1 $\times 10^{-3}$ & 3.0 $\times 10^{-4}$ \\
   $\partial_{\partial_r v_r} \Delta \epsilon$ & 0.41 & 0.2 & -0.08 \\
   
\hline                                   %inserts single line
\end{tabular}
\label{tab1}\end{table}

\subsection{Effects on the Two-Point Correlation Function} 

We calculate the two-point correlation function (2PCF) for our
simulation data following \cite{Landy1993} using a standard estimator
that is frequently written as: 
\begin{equation}
 \xi(r_\perp,\pi) = \frac{DD-2DR+RR}{RR}\,\,,
\end{equation}
where $DD$ is the pair count of galaxies in the simulations, $RR$ is
the pair count of random positions drawn from a uniform distribution,
and $DR$ is the count for pairs consisting of one random location
and one galaxy, all separated by a distance $\pi$ along the line of
sight and $r_\perp$ orthogonal to that. 

\begin{figure*}
 \includegraphics[width=6cm]{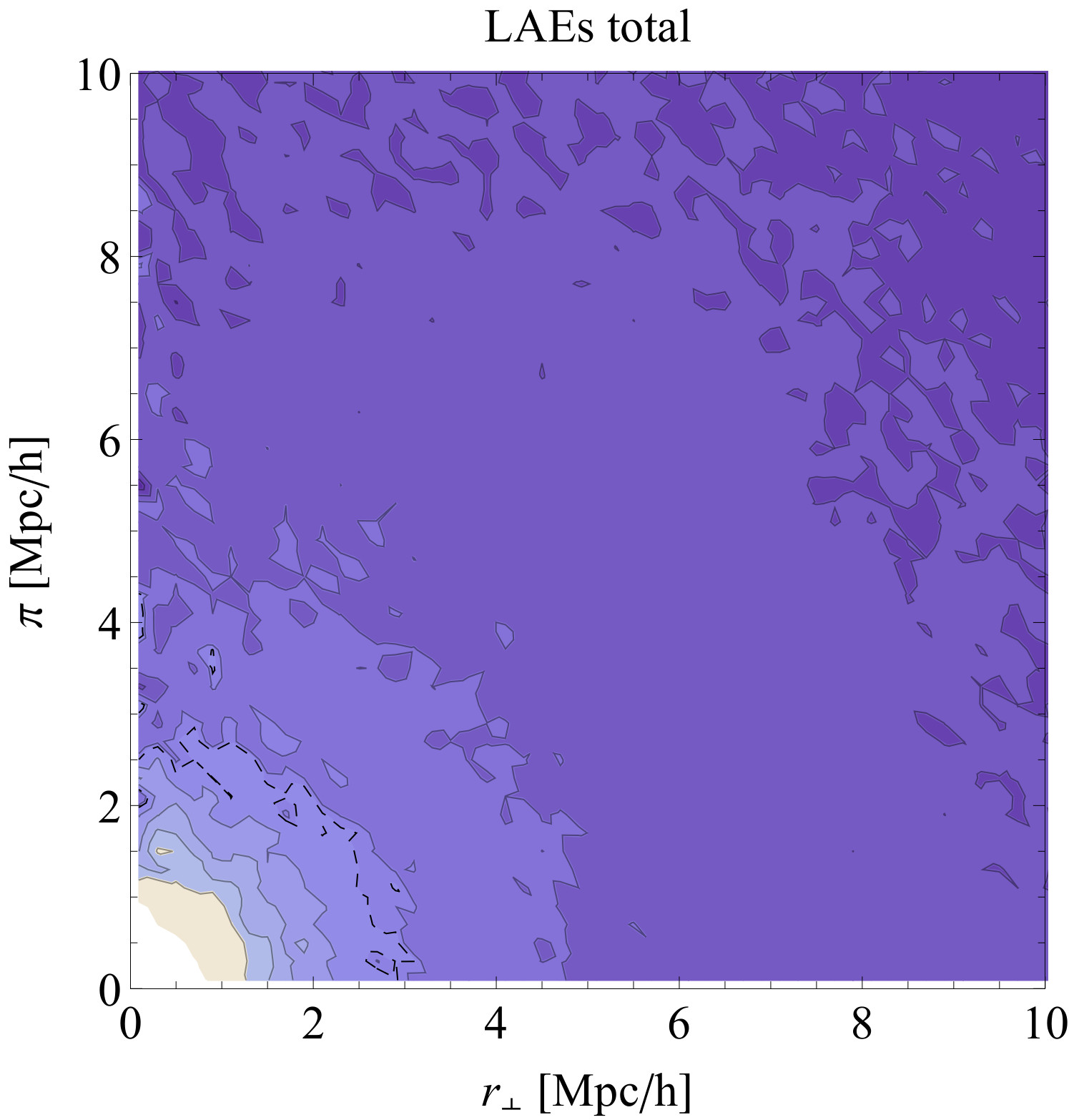} 
 \includegraphics[width=6cm]{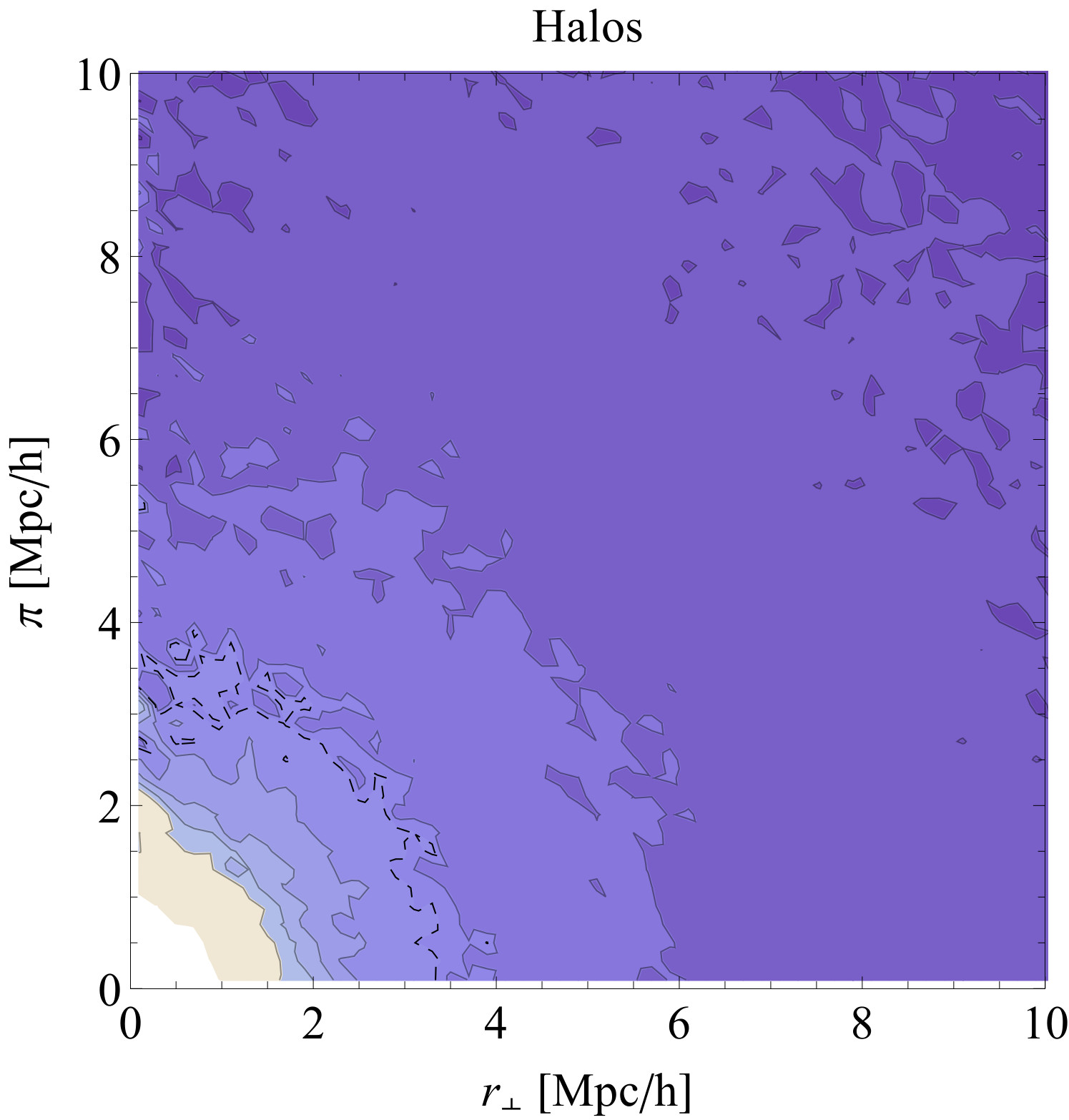} 
 \includegraphics[width=6cm]{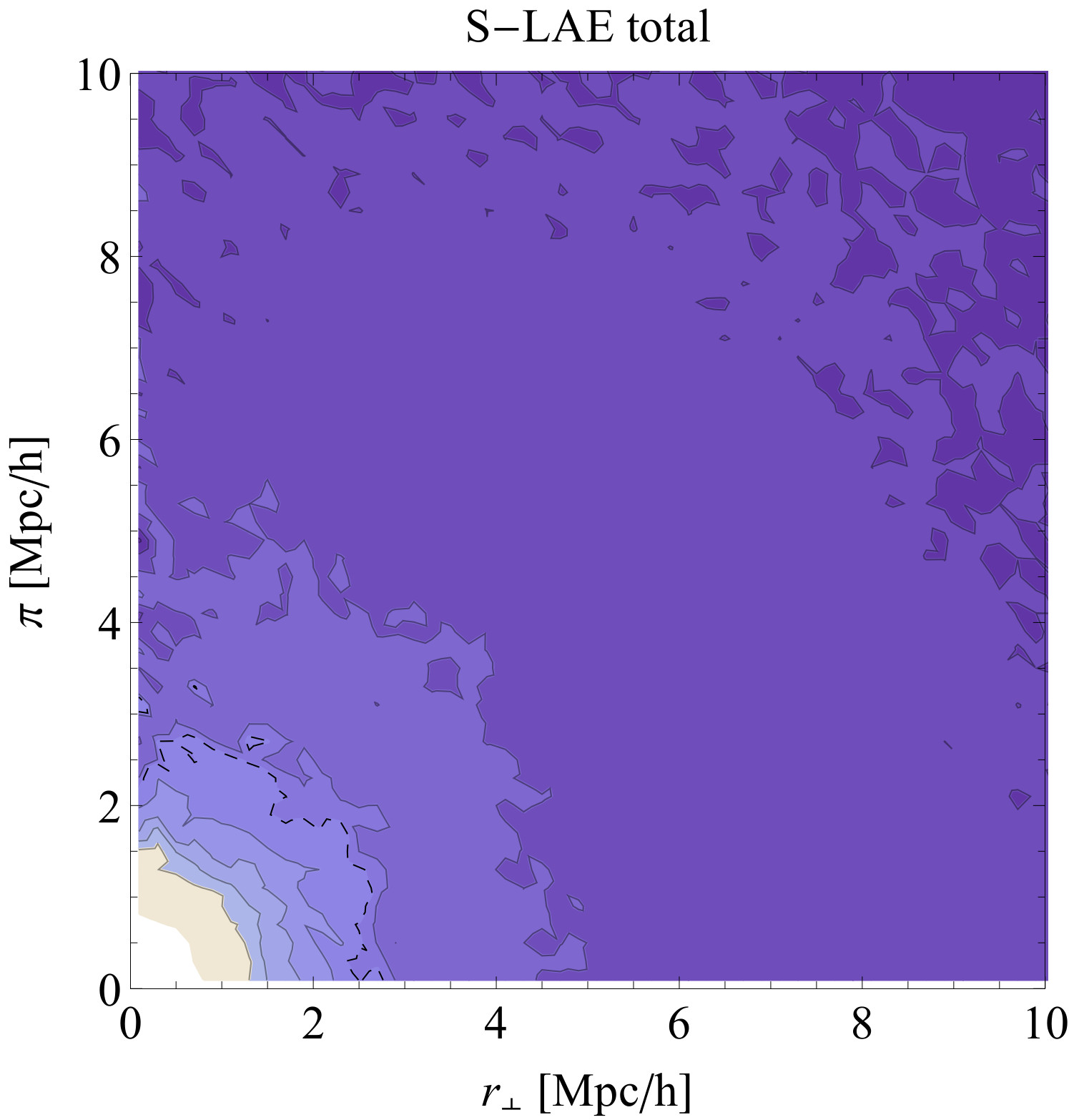} 
 \caption{2PCF as a function of line of sight ($\pi$) and perpendicular distance ($r_\perp$) for halos (left plot), observed LAEs (middle) and S-LAE sample (right) in real space at $z=4$.  The dashed contours corresponds to $\xi=1$, contours are separated by 0.4. The number density of all threshold samples is 4 $\times$ $10^{-2}$ Mpc$^{-3}$ h$^3$. The plots were averaged over three different lines of sight to reduce cosmic variance. }
 \label{fig_2pcf}
\end{figure*}

The real space 3D-2PCF as a function of line of sight separation ($\pi$) and orthogonal
separation ($r_\perp$) is shown in fig. \ref{fig_2pcf} for halos (left), LAEs (middle) and a shuffled LAE sample (right)for our  fiducial case with the \lae detection limit set as
explained above. The shuffled LAE sample was constructed following ZTCM11 by randomly shuffling the properties of the simulated LAEs to get rid off any correlation between apparent luminosity/observed fraction and the spatial location within the box. For all three samples, the number density was fixed to 4 $\times$ $10^{-2}$ Mpc$^{-3}$ h$^3$ which corresponds to a (apparent) luminosity threshold of 0.6 $\times$ $10^{42}$ erg/s for the LAE/S-LAE sample and a mass threshold of 3.6 $\times$ $10^{10}$ $M_\odot$ for the halos. The redshift is $z = 4$ for which we obtained the
largest amplitude in correlations (see above). Near to an orthogonal
separation of $\sim$ 0, the plot shows relatively strong
fluctuations. Those are induced by the small sample size near $r_\perp \sim 0$ and by the finite resolution of our output
grid, resulting in source blending; the projected distance of LAEs in
this region is too small to disentangle them. We do not find a
significant deformation of the 2PCF. 
 ZCTM10 report an strong elongation pattern in the 2PCF along the line of sight
direction which they attribute to a correlation between observed
fraction and line of sight velocity gradient.
In fig. \ref{fig_2pcf_diff}, we plot the relative deviation of
  the LAEs' 2PCF with respect to the shuffled sample, 
$\xi_{\rm diff} = (\xi_{\rm LAE}-\xi_{\rm S-LAE})/\xi_{\rm S-LAE}$. As can be seen, there is no indication of a deformation. We also tried different higher detection limits and luminosity thresholds, but didn't find a significant elongation pattern.  Also, the
correlation with line of sight velocity gradient is not stronger with a higher detection threshold. We conclude that in contrast to ZCTM11,
we do not find a significant elongation in the
line of sight direction.

In addition to the visual inspection of the 2PCF contours, we computed
its quadrupole moment in order to quantitatively verify the absence of
a distortion effect from \lae RT. In fig. \ref{fig_q}, we show the normalized quadrupole $Q(s)$ (e.g. \citet{Chuang2012}) as defined by
\begin{equation}
 Q(s) = \frac{\xi_2(s)}{\xi_0(s) - 3/s^3\int_0^s \xi_0(s')s'^2ds'}
\end{equation}
where $\xi_0(s)$/$\xi_2(s)$ is the monopole/quadrupole contribution as a function of $s = \sqrt{\pi^2+r_\perp^2}$ .

Even for a threshold of $1 \times 10^{-17}$ 
$\textrm{erg s}^{-1} \textrm{cm}^{-2} \textrm{arcsec}^{-2}$
which removes 90 \% of all emitters, no significant signal was found
in the magnitude of the quadrupole moment.

Again, this result can at least in part be attributed to our lower
redshift. The amplitude of the selection effect induced by the RT and
the observation threshold strongly depends on the optical depth in the
IGM. The denser and more neutral IGM in ZCTM10/11 results in a stronger
dimming of the central regions of a source, leading to a lower
observed fraction. By tuning the observation threshold alone, this
cannot be mimicked.  

On the other hand, the linear analysis by \cite{Wyithe2011a}, evaluated with
coefficients estimated from our results, indicates that our box size
may be insufficient to measure a signal. Specifically,
\cite{Wyithe2011a} present an analytical model for estimating the
impact of the radiative transfer on the clustering signal. 
The parameters $C_v$ and $C_\rho$ defined in 
their eq.\ 12 and 13 measure how the relative transmission
of the IGM is affected by fluctuations in the velocity gradient and
density field, respectively. These quantities are  
comparable by construction to 
$\partial_{\partial_r v_r} \Delta \epsilon$ and
$\partial_\delta \Delta \epsilon$
used above. 

We compare with our results in table 
\ref{tab1} by computing $C_v$ and $C_\rho$ with values estimated from
our data, namely the fraction of photons scattered in the IGM,
$F\approx 0.7$, the mean IGM optical depth, $\tau_{IGM}\approx 1.5$, and
the luminosity function power law index, $\beta=2.2$. We obtain 
$C_v \approx 0.6$  and 
$C_\rho \approx -1$ from the analytic model,
which is broadly consistent with our values for $\partial_{\partial_r
  v_r} \Delta \epsilon$ and $\partial_\delta \Delta \epsilon$. At this
level of $C_v$, 
the analysis of \cite{Wyithe2011a} would suggest a
small but noticeable deformation of the 2PCF. We
consequently 
cannot rule out that 
the absence of a signal in our results is affected by
our limited statistics.

\subsection{Impact of Inclination on the Observed Fraction} 

\begin{figure}
   \centering
   \includegraphics[width=\linewidth]{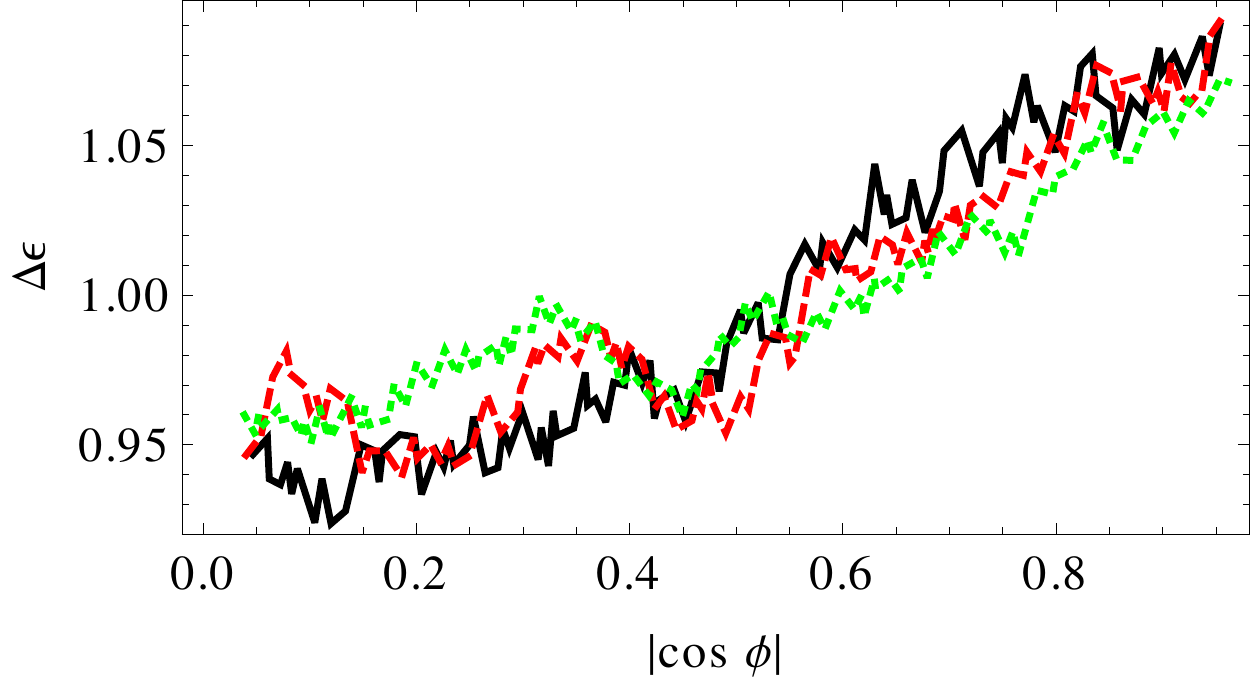}
      \caption{Correlation between the relative
        observed fraction and the inclination of the halos with
        respect to the observer. An inclination of 1 corresponds to a
        face-on emitter, 0 corresponds to edge-on. The lines show data
        from $z=4$ (black), $z=3$ (red) and $z=2$ (green)} 
         \label{fig_inclination}
   \end{figure}

In fig. \ref{fig_inclination}, the correlation between observed
fraction and the inclination angle between the line-of-sight and the angular
momentum of the dark matter halo
is shown for all three redshifts. For this plot, we use the full sample of simulated LAEs. We use the dark matter angular
momentum as a proxy for the disk orientation in order to avoid
ambiguities in the definition of the disk plane, given the marginal
resolution of galactic disks in the MareNostrum simulation.Of course,
the scatter between disk orientation and halo angular momentum
reduces the signal from disk orientation effects on the
\lae observed fraction as seen, for instance, in \cite{Bett2012}. On the other
hand, this choice provides a more direct measure of the impact of halo
tidal alignment on the clustering statistics, to be discussed
below. For simplicity, we will refer to galaxies viewed along the
direction of their halo angular momentum as ``face-on'', in spite of
the fact that the viewing direction may not be exactly normal to the
disk plane.

For face-on galaxies, the observed fraction
is increased by $\sim$15\% with respect to edge-on galaxies. This is
intuitive, since photons preferentially escape perpendicular to
the disk because of the reduced optical depth compared to the path
through the disk plane. Since this is local effect, it
is independent of the chosen line of sight and is not affected by
switching off/on the peculiar velocity field or the Hubble flow.  

Strong inclination dependence has also been found by various groups
\citep[e.g. ][]{Laursen2007b,Yajima2012a,Verhamme2012} in simulations of  isolated disk galaxies. Their
results also showed a strong sensitivity on the morphology of the ISM,
with a denser and clumpier structure exhibiting a significantly more
pronounced dependence on viewing angle. In order to assess the full
extent of LAE emission characteristics as a function of inclination
angle, high-resolution simulations which provide a fair representation
of the ISM morphology are required. Our results, based on a
simulation with marginal spatial resolution of galaxies which results
in a very smooth ISM structure, can therefore only provide a lower
bound on the expected magnitude of the effect. 
 
We do not find a significant evolution of the inclination dependence with
redshift. This is surprising, since one might expect the disk-like
shape to become more prominent at lower redshift due to the higher total
mass in the ISM. Further investigation is needed to resolve this issue, but one
has to keep in mind that comparisons of the signal's amplitude can be
difficult between different redshifts (cf.\ Sec.\ \ref{sec:redshifts}).

The signature of orientation dependence and tidal alignment on
redshift space distortions (RSD) has been analyzed by \cite{Hirata09} who
concludes that the effect is degenerate with gravitationally induced
RSD (i.e., the Kaiser effect, \citealt{KaiserNick1987}) and may amount to several
percent for reasonable assumptions about alignment and inclination
dependence of the observed flux. However, it is easy to see that the
coefficient that measures the orientation dependence (named $\psi$ in
\cite{Hirata09}) can be made much
larger if one assumes a very steep transition from edge-on to face-on
flux, such as the one observed by \cite{Verhamme2012} in their case
``G2''. The fact that the transition seen in our results (fig.\
\ref{fig_inclination}) is rather smooth can be attributed to two effects that have
already been mentioned above: first, the
spread of disk orientations with respect to halo angular momentum
washes out the overall signal, and second, the spatial resolution is
inadequate to capture the full extent of the expected orientation
dependence. While the former is physical and will be present in real
data, the latter is an artifact of our method.  
\begin{figure}[h]
\centering
\includegraphics[width=6cm]{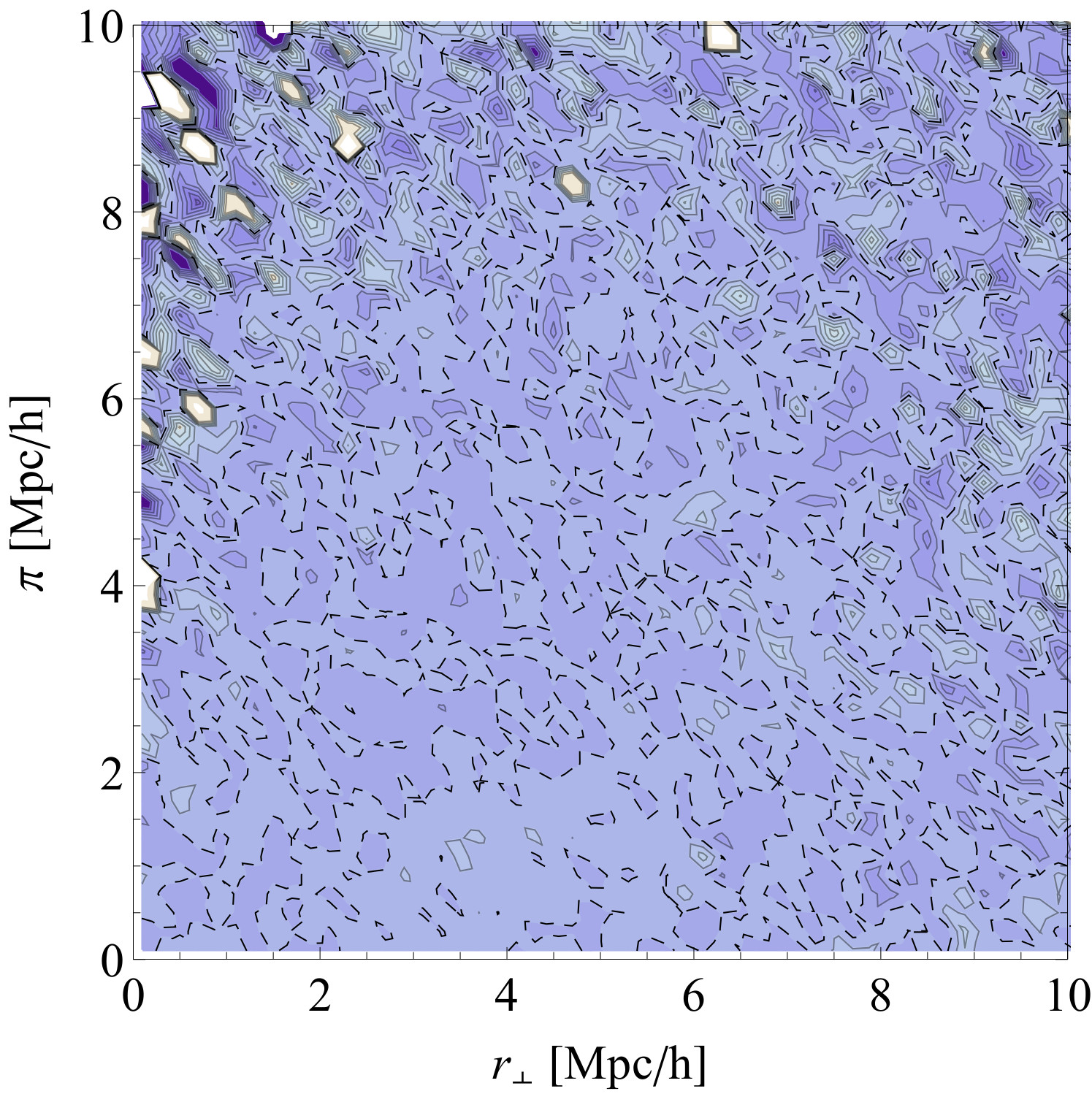} 

\caption{Ratio of the 2PCF of S-LAE and LAEs
  $\xi_{diff} = (\xi_{LAE}-\xi_{S-LAE})/\xi_{S-LAE}$. Contours are separated by 15\%, the dashed contour corresponds to a value of 0. Number densities and thresholds are the same as in fig. \ref{fig_2pcf}.} 
\end{figure}\label{fig_2pcf_diff}

\begin{figure}[h]
\centering
\includegraphics[width=\linewidth]{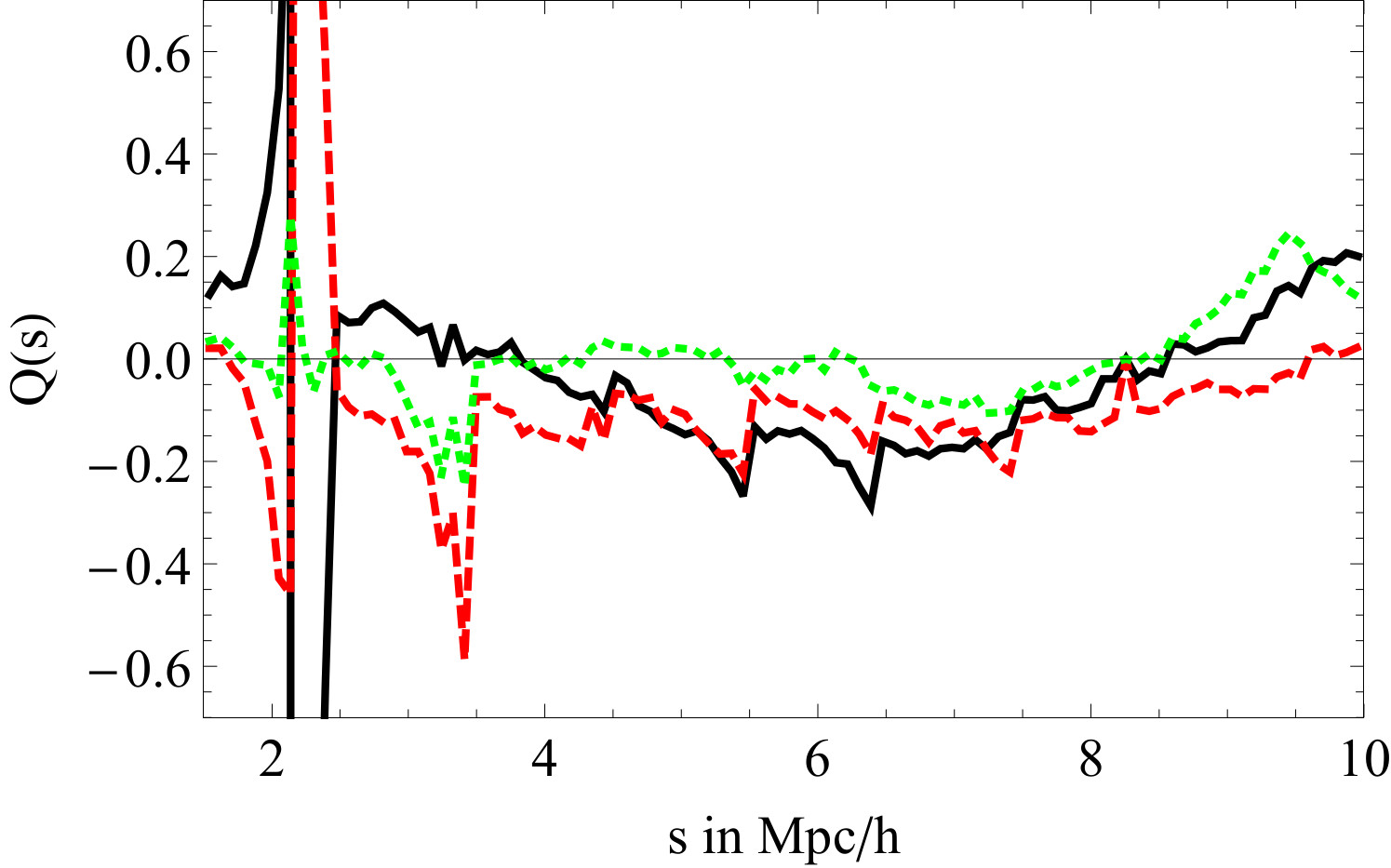} 

\caption{Q(s) for $z=4$ halo sample (black solid), LAE sample (red dashed) and S-LAE sample (green dotted line). Number density is 4 $\times$ $10^{-2}$. The spikes at 2-3 Mpc/h are due to poles.} 
\end{figure}\label{fig_q}

These results are highly suggestive that LAEs can provide a sensitive probe of
gravitationally induced tidal alignment. This could reduce the
accuracy of growth factor measurements from surveys like HETDEX;
however, additional information from the galaxy bispectrum can
break the degeneracy (cf. \citealt{KrauseHirata11}). But it also offers
the attractive opportunity to search for tidal alignment in LAE survey
data and to test the predictions of CDM structure formation. For
instance, recent results from large N-body simulations show an
alignment along large-scale structure filaments for lower-mass halos,
whereas the angular momenta of high-mass halos preferentially align
perpendicular to the direction of filaments
\citep{Codis2012}. Further work is needed to explore the potential of
LAEs as tracers of cosmic alignment.

\section{Conclusions}

Our numerical analysis clearly shows that resonant scattering in the
CGM and IGM can strongly suppress observed \lya fluxes. We find mean
observed fractions between 30\% at $z = 4$ and around 80\% at
$z = 2$. We stress that we do not include destruction by dust, hence the
suppression in flux purely results from anisotropic escape of photons
from their halos and diffuse scattering in the IGM. These results are
consistent with \cite{Laursen2011}.  

We do find correlations between the large-scale density and velocity
fields and the observed \lya fraction.  
This broadly confirms the results of ZCTM10 who report a much stronger effect at higher
redshifts. Apart from their overall smaller amplitude, the
correlations seen in our work differ from those found by ZCTM10 in their
relative strength. Whereas the velocity gradient has by far the
strongest effect in their results, in our case it is comparable in
magnitude to the correlations of \lae observed fraction with large-scale
density, density gradient, and velocity fields. This appears more
natural to us since density and velocity gradient are correlated
via the continuity equation.

All of the correlations with large-scale fields that we found have
plausible interpretations in terms of resonant scattering with neutral
hydrogen modulated by density and Doppler shift. Owing to the strong
correlations between the density and velocity fields in the linear
regime, it is not always possible to unambiguously identify the dominant
effect. By artificially turning off the peculiar velocity and Hubble
flow terms in the scattering cross sections, we were able to separate
the effects of density and velocity to some extent. The results are
consistent with intuitive expectations: the effect of large-scale
overdensity is largest and similar to the velocity gradient, 
followed closely by velocity and density gradient.

For all correlations except orientation dependence, we find a strong
decrease of their amplitude from redshift 4 to 2. Despite the strong
impact of the radiation transport on the observed fluxes, we do not reproduce the
clustering signal found by ZCTM11. Even at $z=4$ where the correlations
are strongest, we fail to detect a significant change in the 3D 2PCF. 

Although the lower redshift of our studies is expected to reduce the
clustering signal, a comparison with the analytical model by
\cite{Wyithe2011a} 
suggests that this is not the full explanation.
The values we find for the dependence of the 
observed flux on the large-scale velocity gradient would, according to
their model, lead to a small but detectable deformation in the 2PCF. 
It is therefore plausible that 
the limited statistics due to our finite box size are partly
responsible for our failure to detect non-gravitational clustering
from \lae RT effects.

We also found a distinctive correlation between the \lae observed
fraction and the angular momentum of the dark matter halo, which we
interpret as a signal of the orientation dependence of LAE
fluxes. Although the amplitude of the signal is only $\sim$ 15 \% in
our numerical analysis, we assume that it can be substantially larger
in reality as our results are limited by poor spatial resolution of the
ISM. In this case, partial alignment of halo spins with the large-scale
tidal field may give rise to contaminating contributions to redshift
space distortions \citep{Hirata09, KrauseHirata11}. On the other hand,
our results combined with recent high-resolution simulations of 
LAEs \citep{Verhamme2012} suggest that LAEs provide a sensitive
observational probe of tidal alignment.

\begin{acknowledgements}
     This work was supported by the 
   DFG SFB 963/1, project A13. We acknowledge the Horizon collaboration for 
making the MareNostrum data set 
available to us. We thank Mark Dijkstra for many fruitful discussions
and comments, and Eichiro Komatsu for helpful suggestions.
\end{acknowledgements}

\bibliographystyle{aa}

\bibliography{library}

\pagebreak
\begin{appendix}

\section{The physics of \lya transport}

\subsection{General}

   \lya photons emitted in starforming regions of a galaxy
   undergo resonant scatterings that result in a stochastic movement
   in space and frequency. As in many previous studies, we denote the
   frequency of a \lya photon with the dimensionless quantity 
\begin{equation}
x = \frac{\nu - \nu_0}{\nu_D}
\end{equation}
where $\nu_D$ is the Doppler frequency, $\nu_D =
\frac{v_{th}\nu_0}{c}$ with the most probable thermal velocity of the atoms
$v_{th}= \left( \frac{2 k_B T}{m_H}\right)^{1/2}$ and all other
symbols having their usual meaning.  

A non-zero bulk velocity of the gas $\vec v$ can be taken into account
easily by performing a first-order Lorentz transformation into the
restframe of the macroscopic gas motion: 
\begin{equation}
x' = x - \frac{\vec v \cdot \vec n}{v_{th}}
\label{eq_lorentz}\end{equation}

Following \cite{Dijkstra2006}, we denote quantities measured in the
restframe of macroscopic bulk velocity with a prime. If not mentioned
otherwise, quantities are measured in the frame of an observer which
is at rest with respect to the center of mass (but notice the remarks
on the Hubble flow in section \ref{section_code}). 
Here, $\vec n$ is the direction of the photon. 

\subsection{Absorption and Reemission}

The scattering cross section of a \lya photon can be written as
\begin{equation}
\sigma_L = f_{12} \frac{\sqrt{\pi}e^2}{m_e c \nu_D} H(a,x')
\end{equation}
with $f_{12}$ the Einstein coefficient and $H(a,x')$ the Voigt profile
that depends on the dampening parameter $a = \frac{\Delta \nu}{2
  \nu_D}$. $\Delta \nu$ is the natural line width. 

Therefore, the optical depth $\tau$ for \lya traveling a
distance $l$ with frequency $x$ through a gas with neutral hydrogen
number density $n_H$ is 
\begin{equation}
\tau = \int_0^l \sigma_L\;n_H dl'
\label{eq_tau}\end{equation}

The probability $P$ of a photon to pass through an optical depth
$\tau$ without being absorbed is equal to  
\begin{equation}
P = e^{-\tau}
\label{eq_tau_exp}\end{equation}

Neutral hydrogen atoms on which the scatterings occur follow a
specific velocity distribution due to their random thermal velocity and
the macroscopic gas velocity, since photons are red/blueshifted in the
frame of the scattering atom. It is convenient to split the thermal
velocity into components parallel and orthogonal to an incoming
photon. The PDF of the parallel component is 
\begin{equation}
P(v_z) = \frac{a e^{-v_z^2}}{\pi [(x'-u_z)^2+a^2]} H^{-1}
\label{eq_pdf}\end{equation}
The other two components orthogonal to the direction of the infalling
photon follow a Gaussian distribution. 

Absorption is quickly followed ($\Delta t \sim 10^{-9}$ s) by
reemission. The frequency of the reemitted photon depends on the
scattering atom's velocity due to the fact that the scattering is
coherent in the atom's restframe, but not necessarily in the reference
frame of an observer. If $\vec v_a$ denotes the atom's velocity in the
frame of the observer, then the relation between the frequency of the
infalling photon $x_i$ and the reemitted photon $x_r$ satisfies  
\begin{equation}
 x_r = x_i + \frac{\vec v_a \cdot (\vec n_r - \vec n_i)}{v_{th}}
\label{eq_reemission}\end{equation}

Where $\vec n_i$/$\vec n_r$ is a unit vector in the direction of the infalling/reemitted photon.
We neglect the recoil on the scattering atom here, because it has been
shown to have no significant effect on the radiation transport
\citep{Zheng2002}. 
The distribution of the remission's direction is determined by a phase
function. Depending on the frequency of the infalling photon,
different phase functions have been proposed for the angular
distribution of \lya photons. As shown in
\cite{Tasitsiomi2005}, for numerical simulations the differences
between the phase function are quickly washed out by the resonant
scatterings. For that reason, we use an isotropic phase function that
can be written as 
\begin{equation}
 P(\vec n_r | \vec n_i) = \textrm{const.}
\label{eq_reemission2}\end{equation}
Since the frequency of the photons generally changes due to the
scatterings, the photons perform a random walk in space and frequency
\citep{HarringtonJ.Patrick1974}. The cross section quickly decreases
when a photon leaves the line center, so the mean free path will
increase drastically for photon left/right of the line
center. Typically, photons leave an optical thick medium after a
couple of scatterings on atoms to which they appear strongly red- or
blueshifted, because in that case, it is probable that the photon is
reemitted with a frequency far away from the line center, measured in
the reference frame of the observer.

\section{LyS - A \lya Simulation Code} 
\label{section_code}

LyS is our implementation of a Monte-Carlo code for \lya
radiation transport. It is capable of tracking photons in grids with
Adaptive Mesh Refinement (AMR). LyS is OpenMP-parallelized, so on
machines with multiple cores, each core can handle one photon at a
time. Additionally, MPI was implemented to deal with the large data set
of the MareNostrum Galaxy Formation Simulation. 
 
Similar to other codes, LyS solves the radiation transfer problem for
individual photons via the following iterative algorithm: 
\begin{enumerate}
 \item Draw an optical depth $\tau_0$ exponentially distributed (eq \ref{eq_tau_exp})
\item Integrate the optical depth $\tau$ while the photon traverses a grid of gas cells (eq \ref{eq_tau})
\item When $\tau$ equals $\tau_0$, a scattering point is reached. Draw
  the velocity of the scattering atom from the PDF in eq. \ref{eq_pdf} 
\item Calculate the new frequency and direction of the scattered
  photon according to eq. \ref{eq_reemission} and \ref{eq_reemission2} 
\end{enumerate}
The iteration is stopped when a photon has traveled a quarter of the
box length from its source. If it it reaches a box boundary before, we
apply periodic boundaries.  

For the generation of the output, we use the so-called next event
estimator or peeling-off method \citep{Whitney2011}. At each
scattering, we calculate the probability that the photon is reemitted
into the direction of the observer and reaches the boundary of the box
without further scatterings. This might be thought as sending out a
'tracer photon' at each scattering event. The probability of reaching
the observer is given by eq. \ref{eq_tau_exp}, where $\tau$ is now the
optical depth along the line of sight from the scattering point to the
boundary of the box. Since the frequency of the reemitted photon
depends on the direction into which it is emitted, one has to assign
the frequency for the tracer photon according to
eq. \ref{eq_reemission}. The calculated probability is summed in the
output array for each scattering. Assuming the observer is located in
the direction of the positive x-axis, the array holds $n_y \times n_z
\times n_\lambda$ bins for the y-/z-coordinate of the photon and the
physical wavelength. 

The line of sight integration used in the peeling-off method also
applies periodic boundaries if the integration distance is less than a
quarter of the box length (and stops the integration if this value is
reached). This removes edge effects due to the finite extent of the
box. One could also just skip sources near to the boundaries, but this
would result in the loss of many emitters, degrading the
statistics. Physically, this corresponds to a flattening of the volume
in the line of sight direction. This is plausible since the box is
thin relative to the distance to the observer. 
 
During the line-of-sight integration, tracer photons are redshifted according to the linear Hubble Law. 

Regular photons are also redshifted on their path through the volume. This is implemented by adding a term 
\begin{equation}
 \vec v_H = -H\vec d_{lsp}\label{eq_hubble}
\end{equation}
to the bulk velocity in eq. \ref{eq_lorentz}, where $\vec d_{scat}$ denotes the distance vector to the last scattering location of the photon and $H$ is the Hubble rate at the specific redshift. In this sense, each photon has its own frame of reference, in rest with respect to the observer's frame, but seeing a spherical velocity field centered on the last scattering location.

To find random numbers following the distribution in eq. \ref{eq_pdf}, we use the so-called rejection method \citep{Press2007} in an implementation similar to \cite{Laursen2007}.
LyS uses the acceleration scheme proposed by \cite{Ahn2002} which reduces the number of scatterings by skipping so-called core scatterings. This is done by cutting off the velocity distribution of the scattering atoms below some value, which effectively forces photons to be scattered by atoms to which they appear far in the blue or red.  If the frequency $|x|$ is below a critical frequency $x_{cw}$, the components of the scatterings atom's velocity that are perpendicular to the photon direction of flight are drawn via
\begin{equation}
 v_{\perp,0} = \sqrt{ x^2_{cw} - \log R_4} \cos(2\pi R_5)
\end{equation}
\begin{equation}
 v_{\perp,1} = \sqrt{ x^2_{cw} - \log R_4} \sin(2\pi R_5)
\end{equation}
Here, the $R_i$ are random numbers drawn from a uniform distribution.

In this paper, we focus on \lya radiation from star forming regions near the center of young galaxies. Taking into account the resolution of our simulation box, it is a good approximation to emit photons at the center of the halos. Following ZCTM10, we choose the intrinsic \lae luminosity of a halo proportional to its star formation rate $R_{SF}$, which is in turn a linear function of the halo mass $M_h$:
\begin{equation}
 L_i = 10^{42} R_{SF} \frac{\textrm{erg} \textrm{s}^{-1}} {\textrm{yr}^{-1} M_\odot}
\end{equation}
\begin{equation}
 R_{SF} = 0.68 \frac{M_h}{10^{10} M_\odot} \textrm{yr}^{-1}
\end{equation}
Both equations should vary with redshift. The cosmic star formation history reached its peak around redshift 2 \citep{Kobayashi2012}, and the intrinsic \lya luminosity should be modified by dust attenuation that is also a function of star formation history. We ignore this here, since from the perspective of our numerical simulation, the total intrinsic luminosity plays only the role of a normalization, especially because we are mostly interested in ratios between intrinsic and apparent luminosities. We also stress that we chose this specific model to be comparable to previous work, not because it is the relation predicted by the MareNostrum simulation. Photons are emitted with a frequency drawn from a Gaussian. Its width is determined by the viral temperature of the halo:
\begin{equation}
 T_{vir} = \frac{G M_h \mu m_H}{3k_Bc}
\end{equation}
Here, $R_{vir}$ denotes the virial radius of the halo and $\mu$ is the mean molecular weight. In this way, we include the velocity distribution of emitters that are gravitationally bound. Photons are emitted from halos with mass $\geq$ 5 $\times$ $10^9 M_\odot$. Since the range of masses and thus intrinsic luminosities is about 3 orders of magnitude, we follow ZCTM10 in applying a weighting procedure for the individual photons to reduce the total number of photons to compute. Each halo emits independently of its mass at least $n_{min} = 1000$ photons. In total, we run the RT for about 40.000/49.000/51.000 halos at redshift 4/3/2. To conserve the relative intrinsic luminosities, photons are given a mass-dependent weight.

\subsection{Code Verification}
\begin{figure}
   \centering
   \includegraphics[width=\linewidth]{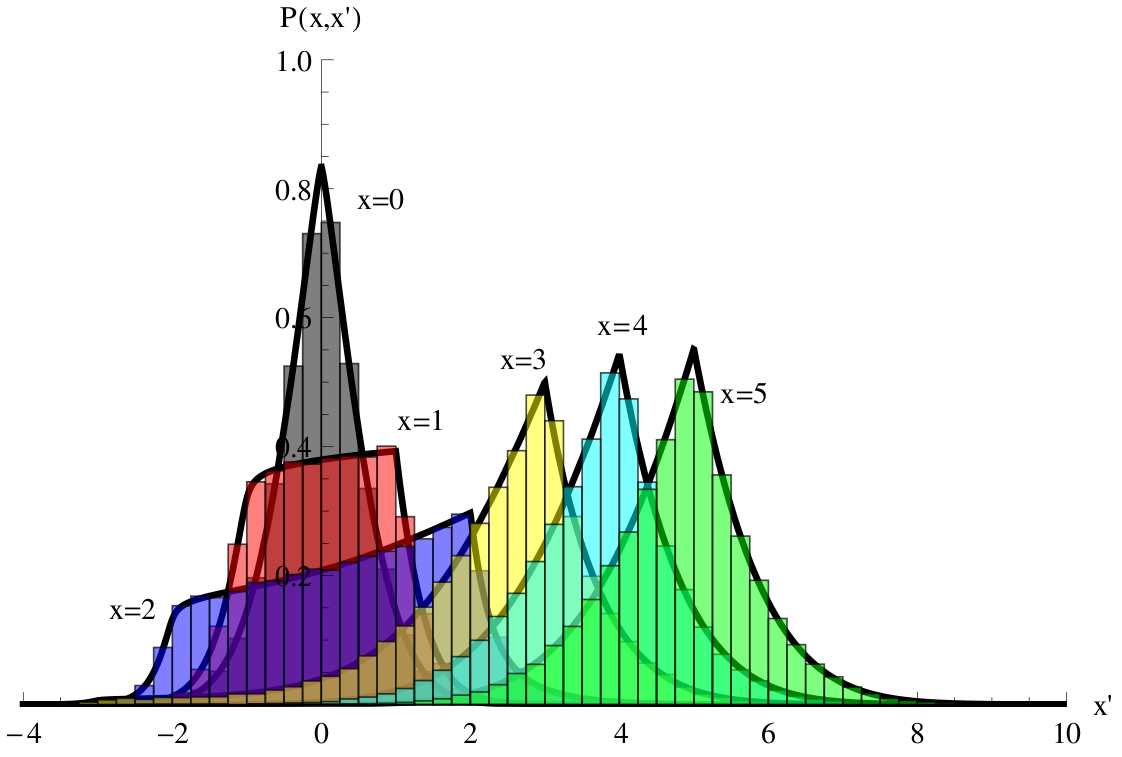}
      \caption{Comparison of the redistribution function calculated using the analytic solution by \cite{Lee1974} (solid lines) with our code. Shown is the probability that a photon is reemitted with a frequency $x'$ when it had the frequency $x$ before the scattering occurred for some values of $x$.
              }
         \label{fig_redist}
   \end{figure}

To verify the correctness of our radiative transfer code, we perform the standard tests from the literature. In figure \ref{fig_redist}, the redistribution function $f(x,x')$ is shown, namely the probability for an infalling photon with frequency $x$ to be reemitted with a frequency $x'$. In this test, only thermal motions of the scattering atom are considered. Overplotted are the analytical solutions by \cite{Lee1974}. 
   \begin{figure}
   \centering
   \includegraphics[width=\linewidth]{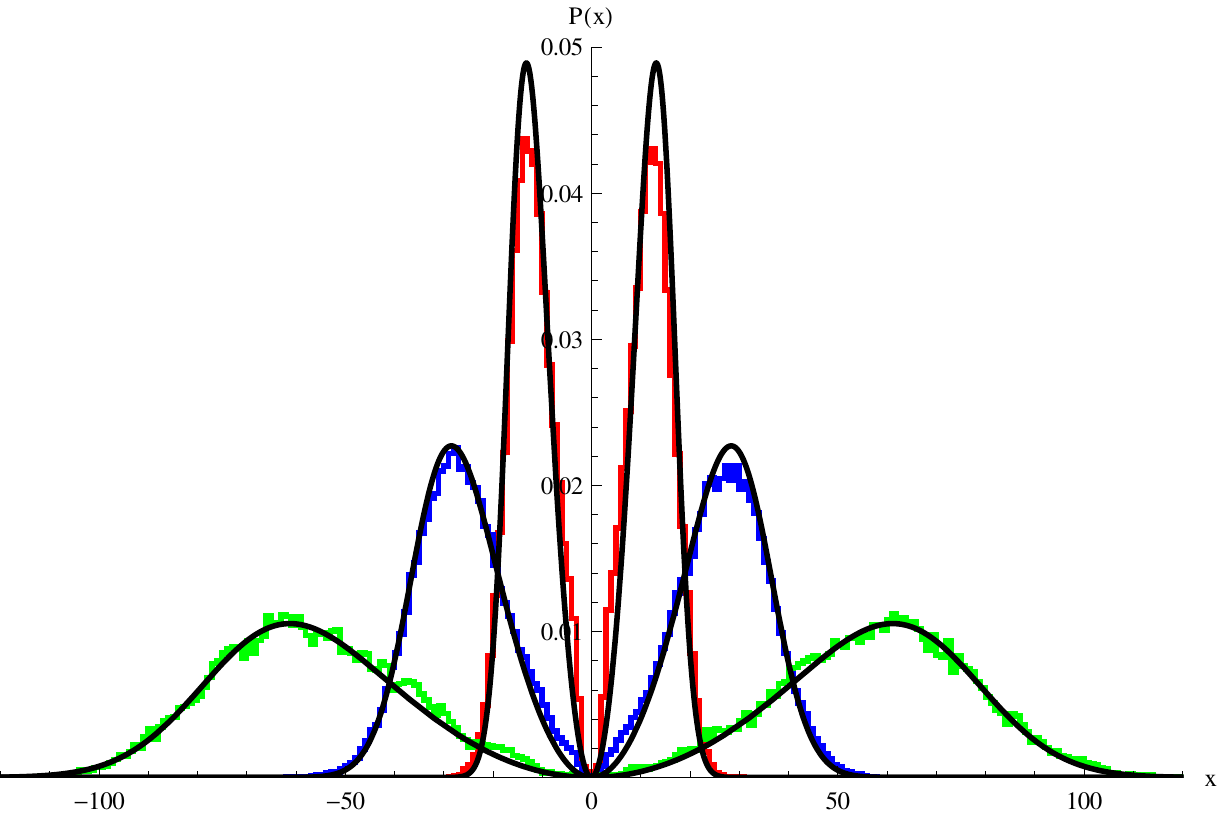}
      \caption{Comparison between the analytical solution (solid lines) of the spherical test case (see text) and the results as obtained by LyS. Shown is the probability distribution of the escaping photons as a function of the dimensionless frequency $x$. The innermost peaks correspond to an optical depth of $10^5$, the outermost to $10^7$. The third case is for an optical depth of $10^6$.
              }
         \label{fig_sphere}
   \end{figure}

The standard test for \lya-Codes, the so-called static sphere test, is shown in fig. \ref{fig_sphere} for various optical depths ($\tau_0=10^5/10^6/10^7$). In this test, photons are launched in the center an isothermal sphere of constant density. Overplotted is the analytic solution from \cite{Dijkstra2006}. For these tests, the acceleration scheme was turned off. The static sphere test was also done with the activated acceleration scheme. It still resembles the analytical solution quite well.

\begin{figure}
   \centering
   \includegraphics[width=\linewidth]{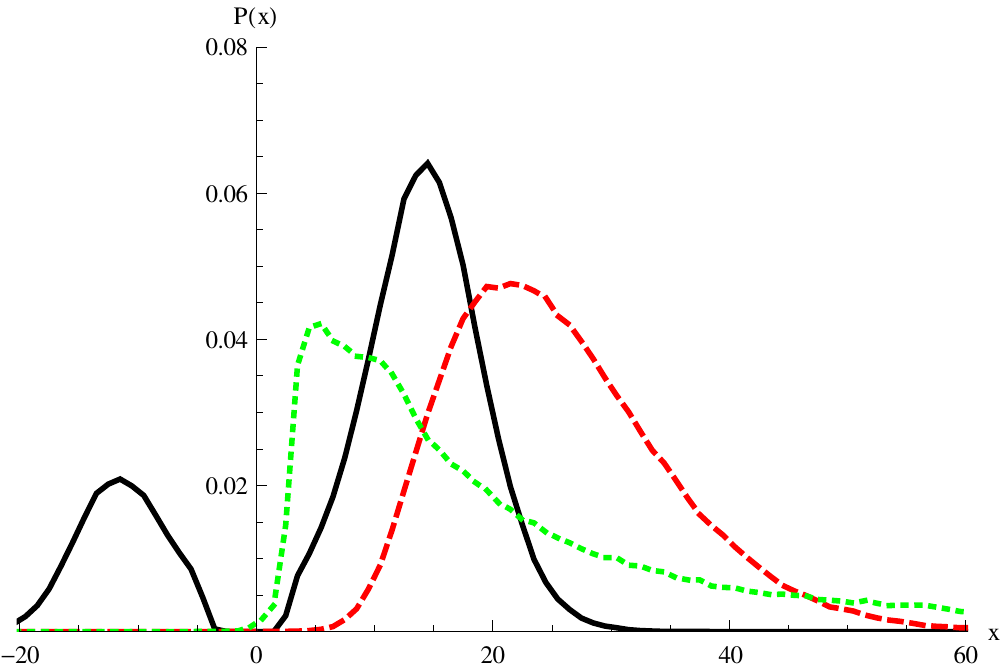}
      \caption{Shown is the dimensionless frequency distribution of photons escaping from an isothermal ($2 \times 10^4$ K) homogeneous sphere with an column density of $2 \times 10^{20}$ $N_H$ from the center to the surface. A Hubble-like velocity prescription given by eq. \ref{eq_velocity} is assigned. The different lines correspond to different maximum collapse velocities: 20 km/s (line), 200 km/s (dashed), 2000 km/s (dotted). 
              }
         \label{fig_sphere_in}
   \end{figure}

\begin{figure}
   \centering
   \includegraphics[width=\linewidth]{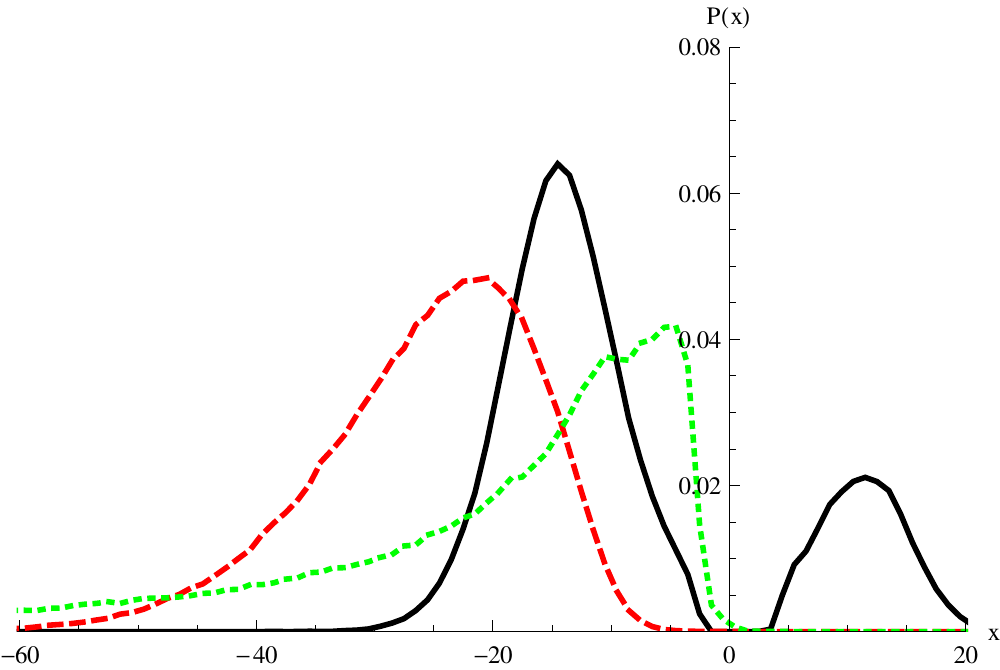}
      \caption{Same as figure \ref{fig_sphere_in}, but for an expanding sphere. The different lines correspond to the maximum expansion velocities: 20 km/s (line), 200 km/s (dashed), 2000 km/s (dotted). 
              }
         \label{fig_sphere_out}
   \end{figure}

For figure \ref{fig_sphere_out} and \ref{fig_sphere_in}, the static sphere problem was modified with a Hubble-like bulk velocity field. The gas was assigned a velocity 
\begin{equation}
 \vec v_{bulk} = \frac{v_{max} \vec r}{r_{max}}
\label{eq_velocity}
\end{equation}
With some constant maximum velocity $v_{max}$, $\vec r$ the distance vector from the center of the sphere and $r_{max}$ the distance from the center where $|\vec v|=v_{max}$. Since there is no analytic solution for this test case available, we can only compare with other the results from other codes. The results from LyS are in good agreement with the plots in \cite{Faucher-Giguere2010a} and \cite{Dijkstra2006}.

\end{appendix}

\end{document}